\documentclass{article}

\usepackage[final]{neurips_2025}

\usepackage[utf8]{inputenc} 
\usepackage[T1]{fontenc}    
\usepackage{graphicx}       
\usepackage{url}            
\usepackage{booktabs}       
\usepackage{amsfonts}       
\usepackage{nicefrac}       
\usepackage{microtype}      
\usepackage{xcolor}         
\usepackage{enumitem}
\usepackage{subfigure}
\usepackage{graphicx}
\usepackage{tablefootnote}
\usepackage{multirow}
\usepackage{multicol} 
\usepackage{algorithm}
\usepackage{algorithmic}

\usepackage{tcolorbox}
\usepackage{xspace}
\usepackage{threeparttable}
\usepackage{url}
\usepackage{titletoc}
\usepackage{wrapfig}
\usepackage{mathtools}
\usepackage{pifont}

\usepackage{bbding}
\usepackage{makecell}
\usepackage{siunitx}

\newcommand{\multirowoffset}{-0.5\dimexpr \aboverulesep + \belowrulesep + \cmidrulewidth}
\newcommand{\model}{UniTraj\xspace}
\newcommand{\datasetname}{WorldTrace\xspace}

\newcommand{\blue}[1]{$_{\color{cyan}\downarrow #1}$}
\newcommand{\red}[1]{$^{\color{red}\uparrow #1}$}

\usepackage[colorlinks,
            linkcolor=cyan,
            anchorcolor=blue,
            citecolor=black
            ]{hyperref}
            
\title{UniTraj: Learning a Universal Trajectory Foundation Model from Billion-Scale Worldwide Traces}

\author{%
  Yuanshao~Zhu\textsuperscript{1,2,3,\thanks{Work done during the internship at HKUST(GZ)}},~~~James~Jianqiao~Yu\textsuperscript{4,\thanks{Corresponding author}},~~~Xiangyu~Zhao\textsuperscript{2,\textsuperscript{$\dag$}},~~Xun~Zhou\textsuperscript{4}, \\ \textbf{Liang~Han\textsuperscript{4},~~~ Xuetao~Wei\textsuperscript{1},~~~
  Yuxuan~Liang\textsuperscript{3,\textsuperscript{$\dag$}}} \\
  \textsuperscript{1}~Southern University of Science and Technology,~
  \textsuperscript{2}~City University of Hong Kong \\
  \textsuperscript{3}~The Hong Kong University of Science and Technology (Guangzhou)\\
  {\textsuperscript{4}~Harbin Institute of Technology, Shenzhen} \\
  \texttt{yuanshao@ieee.org},~~\texttt{jqyu@ieee.org}, ~~\texttt{xianzhao@cityu.edu.hk}\\
  \texttt{zhouxun2023@hit.edu.cn},~~\texttt{han.liang@hit.edu.cn}\\
\texttt{weixt@sustech.edu.cn},~~\texttt{yuxliang@outlook.com}\\
}

\begin{document}

\maketitle

\begin{abstract}
Building a universal trajectory foundation model is a promising solution to address the limitations of existing trajectory modeling approaches, such as task specificity, regional dependency, and data sensitivity.
Despite its potential, data preparation, pre-training strategy development, and architectural design present significant challenges in constructing this model.
Therefore, we introduce \textbf{UniTraj}, a Universal Trajectory foundation model that aims to address these limitations through three key innovations.
First, we construct \textbf{WorldTrace}, an unprecedented dataset of 2.45 million trajectories with billions of GPS points spanning 70 countries, providing the diverse geographic coverage essential for region-independent modeling. 
Second, we develop novel pre-training strategies---Adaptive Trajectory Resampling and Self-supervised Trajectory Masking---that enable robust learning from heterogeneous trajectory data with varying sampling rates and quality. 
Finally, we tailor a flexible model architecture to accommodate a variety of trajectory tasks, effectively capturing complex movement patterns to support broad applicability.
Extensive experiments across multiple tasks and real-world datasets demonstrate that UniTraj consistently outperforms existing methods, exhibiting superior scalability, adaptability, and generalization, with WorldTrace serving as an ideal yet non-exclusive training resource. 
The implementation codes and full dataset are available in the
\textcolor{magenta}{https://github.com/Yasoz/UniTraj}.
\end{abstract}

\section{Introduction}
Trajectory data, as the digital footprints of human movement, is becoming a fundamental data source for understanding mobility patterns and transforming urban intelligence \cite{chen2024deep}. 
These spatio-temporal sequences unlock critical insights across diverse applications: from optimizing transportation networks that alleviate congestion in megacities, enhancing location-based services that personalize user experiences \cite{lan2022vre,chang2023trajectory}, to powering logistics systems that determine the efficiency of global supply chains \cite{guo2018learning, lin2024unite, wang2021survey}. 
Despite their significance, extracting meaningful patterns (from statistical methods to deep learning \cite{luca2021survey}) of trajectory data presents profound challenges due to their inherent complexity, varying lengths, irregular sampling rates, and region-specific characteristics.

As trajectory data continues to expand exponentially, three critical limitations in current approaches have become increasingly apparent:
(1) \textbf{Task Specificity}: Current approaches are typically designed for single-purpose applications, limiting their generalizability and requiring substantial re-engineering for new tasks.
(2) \textbf{Regional Dependency}: Many models are developed and trained on data from specific geographic regions, making them ineffective when applied to different locations with distinct mobility patterns and infrastructure.
(3) \textbf{Data Sensitivity}: Real-world trajectory data often contains noise, irregular sampling, or missing entries, making models highly sensitive to data quality and necessitating extensive preprocessing, which reduces robustness.
These limitations point to a fundamental gap: the absence of a universal foundation model capable of operating across diverse tasks, geographic regions, and data quality levels.
While foundation models have revolutionized NLP \cite{brown2020language,devlin2018bert} and CV \cite{dosovitskiy2020image, he2022masked} by providing versatile, pre-trained architectures that generalize across domains, trajectory analysis has not yet benefited from this paradigm shift. 
Creating this model would transform trajectory intelligence from its current fragmented state to a unified approach with significantly enhanced generalization capabilities \cite{mai2023opportunities,yan2025generative}.

\begin{figure}[t]
    \centering
    \includegraphics[width=\textwidth]{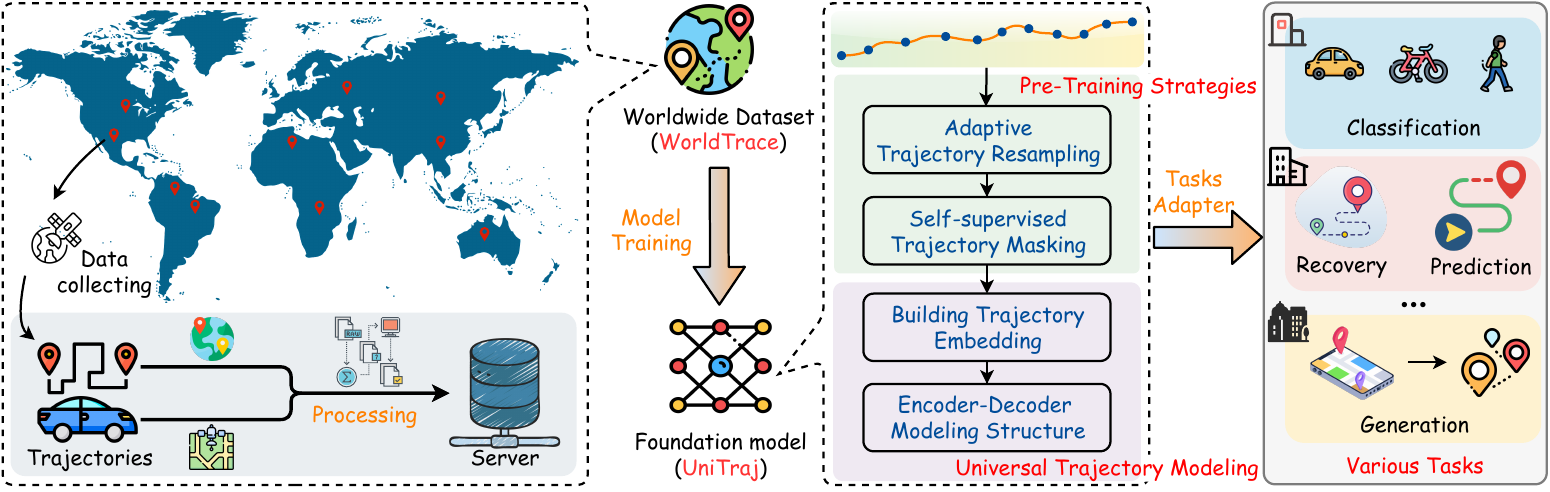}
    \caption{Overview of this work, we propose a trajectory foundation model and also collect a worldwide trajectory dataset. 
    The pre-trained \model can be used as a backbone while adapters are trained for different regions and tasks.}
    \label{fig:intro}
    \vspace{-5mm}
\end{figure}

However, building such a model presents key challenges:
(1) \textbf{Data preparation:} 
The first challenge is to prepare a sufficiently diverse trajectory dataset that spans different geographic regions and appropriate sampling rates.
Existing datasets lack sufficient geographic diversity and scale, also limited by proprietary restrictions and collection costs. This data scarcity severely hampers model generalizability and cross-regional research efforts on a global scale.
(2) \textbf{Pre-training Strategy:}
Developing robust and scalable pre-training strategies is another challenge.
Real-world trajectory data exhibits heterogeneous quality with noising, varying sampling rates, and missing points. Effective pre-training must accommodate these inconsistencies while learning robust representations that transfer across diverse contexts.
(3) \textbf{Model Design:} 
The last challenge involves selecting and tailoring an effective model architecture. 
A universal foundation model requires an architecture that balances adaptability across tasks with computational efficiency, capturing complex spatio-temporal dependencies without overfitting to specific regional information or trajectory  patterns.

To address these challenges, we introduce \textbf{Uni}versal \textbf{Traj}ectory foundation model (\textbf{UniTraj}) supported by three key innovations.
As shown in Figure \ref{fig:intro}, we firstly construct \textbf{WorldTrace}, the first trajectory dataset with large-scale, high-quality, and global distribution, which provides the essential foundation for region-agnostic modeling. 
Then, we design several novel pre-training strategies---adaptive resampling and self-supervised masking---that enable robust learning from heterogeneous trajectory data with varying sampling rates and quality, bridging the gap between regional variations and inconsistent data. 
Finally, we design a flexible model architecture that captures complex spatio-temporal dependencies while adapting to diverse trajectory tasks, creating a versatile backbone for trajectory modeling.
Collectively, UniTraj achieves \textit{task-adaptive, region-independent, and data quality resilience}, delivering a scalable and efficient solution for trajectory analysis applications.
In summary, our research makes the following key contributions:
\begin{itemize}[leftmargin=*]
    \item We introduce WorldTrace, a pioneering trajectory dataset spanning 70 countries with 2.45 million trajectories and billions of GPS points. Its unprecedented global diversity and quality overcome the limitations of existing region-specific datasets, offering a comprehensive and open groundwork for facilitating trajectory modeling research.

    \item We propose UniTraj, trained on WorldTrace and equipped with novel pre-training and masking strategies that effectively capture complex spatio-temporal dependencies. This model significantly enhances generalizability across tasks and geographical contexts, adapts to the heterogeneity of data, and provides a scalable and efficient solution for a wide range of trajectory analysis applications.
    
    \item We demonstrated the effectiveness of UniTraj through comprehensive experiments on multiple trajectory analysis tasks. The results show significantly improved performance of zero-shot and fine-tuning settings, confirming its potential as a versatile backbone for diverse trajectory modeling tasks, performing optimally when trained on diverse and high-quality datasets like WorldTrace. 
\end{itemize}

\section{Related Work}\label{sec:related}

\textbf{Trajectory Datasets}.
Trajectory datasets are foundational for advancing mobility research, yet existing collections vary (geographic coverage, data quality, and granularity) considerably in their utility and limitations.
Well-known datasets, such as GeoLife~\cite{zheng2011geolife}, collected over five years by 182 users, has contributed significantly to fields like travel mode detection~\cite{dabiri2018inferring} and traffic flow analysis~\cite{LI2020225}. However, its limited geographic coverage and participant diversity restrict its generalizability. 
ehicle-focused datasets such as Porto~\cite{Portodata}, T-drive~\cite{T-drive}, and Electric Vehicle Data~\cite{wang2019experience} provide valuable mobility insights but frequently exhibit low or inconsistent sampling rates that complicate analysis. 
Synthetic alternatives like SynMob~\cite{zhu2023synmob} offer uniform sampling but lack the regional diversity and quality variations essential for robust model development. 
Proprietary collections including GAIA~\cite{gaia} and Grab-Posisi~\cite{huang2019grab} contain high-quality data but remain largely inaccessible due to regulatory and commercial constraints.
These limitations---geographic constraints, sampling irregularities, and access restrictions---collectively impede the development of universal trajectory models. 
The community urgently needs comprehensive, openly accessible datasets with global coverage to advance trajectory modeling research and enable effective model generalization.

\textbf{Foundation Models}.
The success of foundation models in natural language processing and computer vision, exemplified by BERT \cite{devlin2018bert}, GPT-3 \cite{brown2020language}, and Vision Transformers \cite{dosovitskiy2020image}, has demonstrated how large-scale pretraining can yield highly generalizable representations across diverse tasks.
This paradigm has recently extended to time series and spatio-temporal domains, with models like TST \cite{zerveas2021transformer}, TimeFM \cite{das2023decoder}, and Moirai \cite{woo2024unified} leveraging Transformer architectures to capture temporal dependencies. 
In spatio-temporal prediction specifically, approaches such as UniST \cite{yuan2024unist}, Opencity \cite{li2024opencity}, and ClimaX \cite{nguyen2023climax} have shown promise in traffic flow and climate modeling, respectively. 
However, these models often remain tailored to specific tasks or regions, limiting their broader applicability.
Trajectory-specific models like TrajGDM \cite{chu2023trajgdm}, BigCity \cite{yu2024bigcity}, and TrajFM \cite{lin2024trajfm} address certain tasks but lack the scalability and robustness needed for cross-task or cross-region applications. 
While unsupervised learning approaches like MAE \cite{he2022masked} and TimeFM \cite{das2023decoder} have proven effective for images and time series, trajectory modeling presents unique challenges that demand greater flexibility to accommodate diverse mobility patterns, geographic contexts, and sampling characteristics without extensive task-specific modifications.
To summary, there remains a pressing need for trajectory foundation models that unify multiple tasks within a single framework, providing robust, transferable representations that generalize across tasks and handle data variability while maintaining computational efficiency. 

\section{Preliminary}\label{sec:pre}

\textbf{Definition 1:} (\textbf{Trajectory}).
A trajectory represents the sequential record of movement through space over time.
Formally, a trajectory $\boldsymbol{\tau}$ of length $n$ is expressed as a sequence of continuously sampled GPS points:
$\boldsymbol{\tau}=\{p_1,~p_2,~\ldots,~p_n\}$, where each point  $p_i = \langle \textnormal{lng}_i, \textnormal{lat}_i, t_i \rangle$ denotes the spatial coordinates (longitude and latitude) at timestamp $t_i$.
The \emph{sampling interval} between consecutive points is defined as $\Delta t_i = t_i-t_{i-1},$ for $i=2,\ldots,n$.
These intervals may be uniform within or across trajectories, or vary significantly based on data collection methods and environmental factors.

\noindent \textbf{Definition 2:} \textbf{(Trajectory Dataset)}.
A trajectory dataset comprises multiple trajectories, each capturing the movement of an object over time.  Formally, it is given by $\mathcal{D} = \{ \boldsymbol{\tau}_1, \boldsymbol{\tau}_2, \dots, \boldsymbol{\tau}_{|\mathcal{D}|} \}$,
where $|\mathcal{D}|$ denotes the total number of trajectories in the dataset.
These collections may vary in geographic coverage, sampling rates, and quality depending on their source and application scenario.

\textbf{Problem Statement:} \textbf{(Universal Trajectory Modeling)}. 
Building upon the above definitions, this study aims to develop a universal foundation model for trajectory data that can adapt to diverse tasks and geographic contexts while accommodating heterogeneous data sources. 
Formally, consider a set of trajectories $\mathcal{D}=\lbrace \boldsymbol{\tau}_i \rbrace_{i=1}^{|\mathcal{D}|}$, where each $\boldsymbol{\tau}_i$ is defined as in Definition 1. 
The goal is:
\begin{align}
    F: \boldsymbol{\tau} \mapsto \mathbf{h} \in  \mathbb{R}^d,
\end{align}
which projects a raw trajectory $\boldsymbol{\tau}$ into a d-dimensional representation $\mathbf{h}$. 
This function $F(\cdot)$must capture intrinsic spatio-temporal patterns within trajectories while demonstrating three key capabilities:
(1) task adaptability across various applications including classification, prediction, and anomaly detection; (2) region independence, enabling zero-shot generalization to different geographic contexts; and (3) resilience to data quality variations, effectively handling inconsistent sampling rates, varying trajectory lengths, and noise without extensive preprocessing or task-specific re-engineering.

\section{Methodology}\label{sec:method}

In this section, we describe the methodology for developing UniTraj, addressing the key challenges outlined in the introduction. 
Our approach is structured around answering three fundamental questions: 
(1) \textit{How to construct a diverse and high-quality trajectory dataset that enables cross-regional generalization}?
(2) \textit{How to develop robust and scalable pre-training strategies that accommodate heterogeneous trajectory data}?
and (3) \textit{How to design an effective model architecture that adapts across diverse trajectory tasks}?

\subsection{WorldTrace Dataset Construction}\label{sec:worldtrace}
To address the data preparation challenge, we introduce WorldTrace, a large-scale, globally distributed trajectory dataset specifically designed to support universal trajectory modeling.
Below, we introduce our data acquisition process, preprocessing pipeline, and key dataset statistics, demonstrating WorldTrace's suitability as a foundation for developing robust and generalizable trajectory foundation models.
Detailed information on processing, analysis, and copyright can be found in \textbf{Appendix \ref{app:dataset}}.
The full dataset is available on the Hugging Face\footnote{https://huggingface.co/datasets/OpenTrace/WorldTrace} and ModelScope\footnote{https://www.modelscope.cn/datasets/opentrace/WorldTrace} platforms.

\begin{wraptable}{r}{6.5cm}
    \fontsize{9pt}{12pt}\selectfont{\caption{Summary statistics of WorldTrace.}\label{tab:data_stats}}
      \fontsize{9pt}{12pt}\selectfont
    \begin{tabular}{lrr}
    \toprule
    \textbf{Statistic} & & \textbf{Value} \\
    \midrule
    Number of Trajectories & & 2.45 Million \\
    Total Raw Points & & 8.8 Billion \\
    Geographical Covered & & 70 Countries \\
    Sampling Interval & & \SI{1}{\sec} (normalized) \\
    Time Span & & 08/2021 -- 12/2023 \\
    Avg. Duration & & \SI{6}{\min } \\
    Avg. Distance  & & \SI{5.73}{\kilo \meter} \\
    Avg. Speed & & \SI{48.0}{\kilo \meter /\hour} \\
    \bottomrule
    \end{tabular}
\vspace{-12pt}
\end{wraptable}

\textbf{Data Acquisition}.
We sourced raw trajectory data from OpenStreetMap (OSM) GPS traces \cite{OpenStreetMap}, focusing on contributions uploaded between 2021-2023 and tagged for motorized movement to ensure data currency and relevance. 
This approach minimizes device heterogeneity and outdated data impacts. All collected data is stored in the standardized GPX format (an XML schema), containing latitude, longitude, timestamps, and optional metadata, providing a uniform structure that simplifies parsing and preprocessing.
During acquisition, we implemented preliminary filtering to exclude trajectories with obvious anomalies such as coordinates outside valid ranges or duplicate entries.

\textbf{Data Preprocessing}.
Our preprocessing pipeline balances preserving authentic movement patterns with removing noise and inconsistencies, which includes the following steps:
\begin{enumerate}[leftmargin=*] 
    \item \textbf{Normalization}: The original data had a high sampling frequency of up to 10 Hz, causing redundancy and increased storage demands. 
    We therefore resampled trajectories to a uniform rate of one point per second (1 Hz), preserving essential motion details while reducing data size.
    In addition, by standardizing trajectories to 1s/point, we can perform better resampling during subsequent model training to accommodate frequency inconsist issues.

    \item \textbf{Filtering}: 
    We discarded trajectories with fewer than 32 points or covering distances below 100 meters, as such short trips often lack meaningful patterns and introduce noise. 
    Following established practices~\cite{dabiri2018inferring}, we also removed trajectories containing implausible speeds (e.g., exceeding 120 km/h), typically caused by GPS errors or anomalies.
    We also apply distance- and loop-based outlier detection to identify and remove trajectories that deviate markedly from the expected path.

    \item \textbf{Calibration}: 
    Given that GPS signals can suffer from errors due to building obstructions, multipath effects, and receiver noise~\cite{hofmann2007gnss}, we applied map-matching techniques~\cite{Yang2018FastMM} to align raw GPS points with underlying road networks.
    This calibration step is common practice in trajectory data processing and is widely used in data collection and related research to correct positioning errors~\cite{gaia,WDR}, improve spatial accuracy, and make trajectory analysis more reliable.
\end{enumerate}

\textbf{Data Analysis and Statistics}.
After acquiring and preprocessing the raw trajectory data, we conducted an in-depth analysis to examine the characteristics and quality of the \datasetname dataset. 
Table \ref{tab:data_stats} summarizes key statistics of \datasetname. 
\textbf{}Overall, the dataset contains approximately 2.45 million trajectories and 8.8 billion raw GPS points, covering 70 countries across all inhabited continents. 
The data spans August 2021 to December 2023, with an average trajectory duration of about six minutes (with normalized to a 1-second sampling interval),  an average distance of 5.73 km, and an average speed of 48.0 km/h. 
The number of points per trajectory ranges from 32 to more than 600, averaging around 358 points. 
Collectively, these attributes confirm \datasetname's suitability for developing universal trajectory models that can address varied spatiotemporal patterns and broad geographical contexts.

\subsection{Pre-Training Strategies}\label{sec:pre-training}
Having established a diverse trajectory dataset, we develop robust pre-training strategies to learn  robust and transferable spatio-temporal representations. 
Rather than relying on task-specific supervision, we leverage unannotated trajectory data to capture both local and global movement patterns. 
To address the heterogeneous data quality challenges (varying sampling rates, differing lengths, and missing points) posed by real-world trajectory, we propose two strategies tailored specifically for trajectory: \textit{Adaptive Trajectory Resampling} and \textit{Self-supervised Trajectory Masking}. 
Due to space limitations, more details and analysis about pre-training strategies can be found in 
\textbf{Appendix}~\ref{app:pretraining}.

\textbf{Adaptive Trajectory Resampling (ATR)}.
Real-world trajectory data often exhibits inconsistent sampling intervals and lengths due to diverse collection standards, device capabilities, and user behaviors. 
Such discrepancies challenge model generalization, as features learned under one sampling regime may not transfer to another. 
Inspired by common practice of multi-scale representation learning, 
ATR strategy addresses these issues through two complementary components:
\begin{itemize}[leftmargin=*]
    \item \textbf{Dynamic Multi-Scale Resampling}.
    This approach dynamically adjusts sampling frequency based on trajectory length, ensuring shorter trajectories retain fine-grained detail while longer ones are efficiently compressed.
    Specifically, we design a logarithmic resampling function $R(n)$ to implement this strategy:
    \begin{align}\label{eq:sampling}
    R(n) = R_{\text{min}} + (1 - R_{\text{min}}) \cdot \frac{\ln(n - n_{\text{min}} + 1)}{\ln(n_{\text{max}} - n_{\text{min}} + 1)},
    \end{align}
    where $n_{\text{min}}$ and $n_{\text{max}}$ define thresholds for trajectory lengths considered ``short'' or ``long'', and $R_{\text{min}}$ is the minimum sampling ratio.
    This logarithmic function creates a smooth transition in sampling density ($n_{\text{min}} < n < n_{\text{max}}$), providing three key benefits: (1) preserving critical motion patterns across trajectory lengths, (2) reducing overfitting by limiting redundancy in densely sampled data, and (3) exposing the model to diverse temporal resolutions during training.
    
    \item \textbf{Interval Consistent Resampling}. 
    This component focuses on the sampling rate, imposing a uniform time interval $\Delta t$ between consecutive points within each track:
    \begin{align}
    \boldsymbol{\tau}^{\prime} = \{\,p_{k_j} \mid k_j = 1 + (j-1)\Delta t,\; j = 1, 2, \ldots, m\}.
    \end{align}
    By ensuring consistent spacing, this approach simplifies downstream modeling by creating regular temporal structures that make time-dependent patterns easier to learn, while mitigating complications from missing data or irregular sampling.
\end{itemize}
\vspace{-2mm}
Combining these approaches, ATR enables models to learn representations that generalize across varying sampling rates and trajectory lengths (analysis presented in \textbf{Appendix \ref{app:resampling}}), which is a critical capability for universal trajectory modeling.

\textbf{Self-supervised Trajectory Masking (STM)}.
Trajectory data is often incomplete or irregular due to device limitations, communication failures, and environmental factors. 
Motivated by masked auto-encoding methods from visual and language models, we introduce a tailored self-supervised trajectory masking strategy, in which part of the input trajectory is hidden, forcing the model to infer local and global dependencies.
Given a resampled trajectory $\boldsymbol{\tau}^{\prime} = \{ p_1, p_2, \dots, p_n \}$, we define a masking function $\mathcal{M}(\boldsymbol{\tau}^{\prime}, r)$ that replaces a fraction $r$ of points with a [MASK] tokens:
\begin{align}
\tilde{\boldsymbol{\tau}} = \mathcal{M}\bigl(\boldsymbol{\tau}^{\prime}, r\bigr)
= \{p_1, \dots, [\text{MASK}]_{i \in \mathbf{I}}, \dots, p_n \},
\end{align}
where $\mathbf{I} \subseteq \{1,2,\ldots,n\}$ and $r = | \mathbf{I} | / n$. To comprehensively address different data incompleteness scenarios, (see \textbf{Appendix \ref{app:masking}}  for details) we employ four complementary masking strategies:
\begin{itemize}[leftmargin=*]

    \item \textbf{Random Masking:} 
    Uniformly samples points to mask ($\mathbf{I}_{\text{rand}} \sim \text{Uniform}(\{1,2,\ldots,n\})$), forcing the model to infer both short-range and long-range dependencies.
    By forcing the reconstruction of randomly omitted points, the approach enhances the model’s ability to generalize to diverse gaps.
    
    \item \textbf{Block Masking:}
    Conceals consecutive points $(\mathbf{I}_{\text{block}}  =\{k, k+1,\dots,k+b-1 \})$ to simulate sensor failures, encouraging reconstruction of continuous segments.
    This approach prompts the model to utilize surrounding context for reconstructing entire missing segments, encouraging it to capture longer-range dependencies.
    
    \item \textbf{Key Points Masking:}
    Identifies and masks critical turning points using the Ramer-Douglas-Peucker algorithm~\cite{douglas1973algorithms}: $\mathbf{I}_{\text{key}} = \{p_k~|~d_{\text{max}}(p_{k}, \overline{p_1 p_n}) >  \epsilon \}$ ($d_{\text{max}}(\cdot)$ is the maximum perpendicular distance between point $p_k$ and line $\overline{p_1 p_n}$, $\epsilon$ is the threshold). 
    This focuses learning on structurally significant points (sharp turns or notable speed changes) that define the trajectory's shape.
    
    \item \textbf{Last N Masking:}
    Masks final trajectory points ($\mathbf{I}_{\text{last}} =\{n-N+1, n-N+2,\ldots, n\}$).
    This setting emulates real-world forecasting tasks where future data is unavailable and must be inferred from historical observations, making it particularly effective for prediction scenarios.
\end{itemize}
\vspace{-3mm}

\subsection{Universal Trajectory Modeling}
To effectively leverage the diverse trajectory data and robust pre-training strategies described above, we need to design a model architecture that can capture local and global patterns while freeing itself from regional and task-specific constraints.
Our motivation for adopting this structure design is as follows:
(1) We need an architecture that can be generalized to a wide range of tasks without extensive restructuring. Therefore, we adopted minimal trajectory data information (latitude, longitude, and timestamp) and ignored other region-bound information such as POI and geographical context.
(2) This structure uses the reconstruction of missing points in partial observations as a proxy task and can inherit the masking strategy introduced earlier.
(3) The separation of encoding and decoding enables flexible application to various downstream tasks through transfer learning or fine-tuning. 
More details about the architecture and parameters can be found in \textbf{Appendix \ref{app:architecture}}.

\textbf{Building Trajectory Embedding}.
Effective trajectory modeling requires transforming raw spatial and temporal data into structured embeddings that capture both local and global movement patterns. 
To ensure the generality of the model, we only use the latitude, longitude, and time information of the trajectory, and embed the spatial and temporal components separately to form a unified representation.
For the spatial component, we normalize trajectory and map them into a $d$-dimensional space using a 1D convolutional, yielding a spatial embedding $\boldsymbol{h}_i^{\text{s}}$. 
Similarly, the temporal component, based on the time intervals $\Delta t_i$, is embedded into the same $d$-dimensional space via a linear layer, resulting in a temporal embedding $\boldsymbol{h}_i^{\text{t}}$. 
This decoupled design enables the model to effectively learn relative movement and temporal dependencies, and also cope with situations where one component may be absent.
Beyond point-wise embedding, modeling the relationships between trajectory points is critical for understanding movement patterns. 
We adopt Rotary Position Encoding (RoPE)~\cite{su2024roformer}, which applies rotational transformations in the embedding space.
The advantage of RoPE is its ability to preserve relative positional relationships while allowing for flexible encoding of spatial-temporal patterns across varying trajectory scales.

\textbf{Adaptive Representation Learning}.
Based on the trajectory embeddings, we use a encoder-decoder architecture with RoPE-enhanced attention mechanism to adaptively learn a general representation of trajectories.
The encoder processes the visible points in a trajectory those that are unmasked during training. 
Given a masked trajectory $\tilde{\boldsymbol{\tau}} = \{ p_1, \dots, \text{[MASK]}_{i \in \mathbf{I}}, \dots, p_n \}$, we first extract the embedding representations of the unmasked points $\mathbf{H} = \{ \boldsymbol{h}_1, \boldsymbol{h}_2, \dots, \boldsymbol{h}_m \}$ (where $m \leq n$ and $i \notin \mathbf{I}$) through the embedding steps. 
The encoder, denoted as $\mathbf{E_{\theta}}$, processes these visible embeddings to generate latent representations: $\boldsymbol{z}_{\text{enc}} = \mathbf{E_{\theta}}\left( \mathbf{H}  \right)$.
The decoder reconstructs masked trajectory points based on the latent embeddings produced by the encoder.
It receives the visible embeddings and mask tokens, which are initialized as learnable vectors representing missing positions. 
The full sequence is created by merging the encoded visible embeddings with the mask tokens, preserving the original structure of the trajectory:
\begin{align}
    \mathbf{z}_{\text{dec}} = \text{Reorder}\left( \left\{ \begin{array}{ll}
\boldsymbol{z}_i = \boldsymbol{z}_{\text{enc}, j} & \text{if } i = \text{Index}(j), \; i \notin \mathbf{I} \\
\text{[MASK]} & \text{if } i \in \mathbf{I}
\end{array} \right\} \right),
\end{align}
where $\boldsymbol{z}_{\text{enc}, j}$ corresponds to the $j$-th encoder output. 
The decoder then processes the reordered sequence to predict the missing trajectory points: $\hat{\boldsymbol{\tau}} = \text{Linear}(\mathbf{D_{\phi}}(\mathbf{z}_{\text{dec}}))$.
The model is trained to minimize the reconstruction loss between the predicted and original points at the masked positions:
\begin{align}
    \mathcal{L} = \frac{1}{|\mathbf{I}|} \sum_{i \in \mathbf{I}} \left\| f_{\theta, \phi}(\tilde{\boldsymbol{\tau}})_i - \boldsymbol{\tau}_i \right\|^2,
\end{align}
where $f_{\theta, \phi}(\tilde{\boldsymbol{\tau}})$ represents the encoder-decoder network, and $i$ refers to the masked positions.

\section{Experiments}\label{sec:exper}

\subsection{Experimental Setups}
\textbf{Datasets}.
We evaluate UniTraj on six diverse real-world trajectory datasets representing different collection scenarios, quality levels, motion patterns, and geographic regions. These include WorldTrace, Chengdu, Xi'an, GeoLife, Grab-Posisi, and Porto. Detailed summary are provided in \textbf{Appendix \ref{app:exp_dateset}}.

\subsection{Task Applicability Analysis}
We explore the applicability and generalizability of UniTraj to various data and downstream tasks, e.g., trajectory recovery, prediction, classification, and generation tasks. Due to space constraints, we provide the detailed setup and the results of generation task in \textbf{Appendix \ref{app:task_setting}}.
It is important to clarify that our work aims to develop a general-purpose trajectory foundation model that generalizes across diverse geographic regions without region-specific dependencies, validating its effectiveness as a backbone supporting real-world trajectory applications across geographical contexts.
Existing trajectory representation learning methods inherently rely on region-bound information (POIs, road networks, etc.)\cite{jiang2023self,lin2023pre,ma2024more,zhou2024red}, which contradicts our initial goal of region-independent modeling. UniTraj extracts meaningful representations solely from trajectory points without requiring auxiliary geographic context. Therefore, we deliberately excluded these methods from our baseline comparison as their architectural dependency on regional knowledge fundamentally diverges from our objective of developing a globally deployable model.

\begin{table*}[t]
    \caption{\fontsize{9pt}{12pt}\selectfont{Performance comparison of \model with trajectory recovery tasks. The results are reported in MAE and RMSE with meters.
    Bold denotes the best results and \underline{underline} denotes the second-best results. }}
    \centering
    \resizebox{0.99\textwidth}{!}{
    \begin{tabular}{lcccccccccccc} 
    \toprule
      \multirow{2}{*}[\multirowoffset]{\textbf{Methods}}  & \multicolumn{2}{c}{\textbf{WorldTrace}} & \multicolumn{2}{c}{\textbf{Chengdu}} & \multicolumn{2}{c}{\textbf{Xi'an}} & \multicolumn{2}{c}{\textbf{GeoLife}} & \multicolumn{2}{c}{\textbf{Grab-Posisi}}  & \multicolumn{2}{c}{\textbf{Porto}} \\
    \cmidrule(lr){2-3}\cmidrule(lr){4-5}\cmidrule(lr){6-7}\cmidrule(lr){8-9}\cmidrule(lr){10-11}\cmidrule(lr){12-13}
 & \textbf{MAE}  & \textbf{RMSE}  & \textbf{MAE}  & \textbf{RMSE}  & \textbf{MAE}  & \textbf{RMSE}  & \textbf{MAE}  & \textbf{RMSE}  & \textbf{MAE}  & \textbf{RMSE}   & \textbf{MAE}  & \textbf{RMSE} \\ 
    \cmidrule(lr){1-13}
    Linear    & 427.68 & 516.15 & 205.74 & 258.52 & 176.49 & 220.87 & 196.85 & 249.76 & 507.41 & 617.28 & 396.61 & 482.39 \\
    DHTR  & 220.35 & 302.47 & 75.19 & 98.68 & 62.85 & 83.43 & 80.04 & 168.25 & 351.20 & 415.16 & 194.37 & 232.59\\
    Transformer & 130.82 & 147.62 & 55.23 & 62.85 & 45.85 & 51.96 & 94.68 & 113.77 & 136.58 & 163.29 & 104.36 & 126.96 \\
    DeepMove  & 51.16 & 62.29 & 29.32 & 39.02 & 27.31 & 35.67 & 86.38 & 107.78 & 126.93 & 168.07 & 136.66 & 174.96 \\
    TrajBERT   & 58.13 & 70.14 & 26.48 & 33.83 & 19.45 & 25.13 & \underline{34.53} & \underline{43.24} & 112.68 & 136.24 & 78.77 & 99.23     \\
    TrajFM   & 47.64 & 58.92 & 19.10 & 25.09  & 18.86  & 24.13 & 59.34 & 64.24 & \underline{107.64} & \underline{130.69} & \underline{71.15} & \underline{92.96} \\
    \cmidrule(lr){1-13}
    \model (zero-shot) & \underline{10.22}& \underline{13.56} & \underline{11.98} & \underline{20.94}  & \underline{8.93} & \underline{13.83} & 37.21 & 63.89 & 114.07 & 167.01 & 78.28 & 100.14 \\
    Improvement(\%) & \red{78.55} & \red{76.99} & \red{37.28} & \red{16.54} & \red{52.65} & \red{42.69} & \blue{7.76} & \blue{47.46} & \blue{5.97} & \blue{27.79} & \blue{10.02} & \blue{7.72} \\
    \model (fine-tune) & \textbf{6.94} & \textbf{9.67} & \textbf{6.92} & \textbf{10.41}  & \textbf{6.50} & \textbf{9.93} & \textbf{23.23} & \textbf{34.70} & \textbf{48.95} & \textbf{69.23} & \textbf{60.18} & \textbf{79.76} \\
    Improvement(\%) & \red{85.43} & \red{83.59} & \red{63.77} & \red{58.51} & \red{65.54} & \red{58.85} & \red{32.73} & \red{19.75} & \red{54.52} & \red{47.03} & \red{15.42} & \red{14.20} \\
    \bottomrule
    \end{tabular}
    }
    \vspace{-3mm}
    \label{tab:recovery_comp}
\end{table*}

\textbf{Trajectory Recovery}.
Table~\ref{tab:recovery_comp}  presents a comprehensive comparison of UniTraj against established baselines across six datasets, revealing patterns that illuminate fundamental capabilities in trajectory reconstruction.
The performance disparity between UniTraj and previous methods is particularly pronounced in geographically diverse and quality-variable datasets, where it demonstrates substantial resilience to regional variations.
In the zero-shot setting, \model achieves remarkable results, confirming it effectively captures transferable spatio-temporal patterns without requiring additional fine-tuning.
The performance difference becomes particularly instructive when analyzing low-quality datasets like GeoLife and Grab-Posisi, with their highly irregular sampling intervals and multiple travel modes. 
It demonstrates the effectiveness of our adaptive resampling strategy in handling temporal heterogeneity.
The Chengdu and Xi'an datasets reveal another critical aspect of UniTraj's capabilities, models trained on high-quality data exhibit reliable transferability and achieve optimal results even in zero-shot scenarios.
When fine-tuned, \model achieves the lowest error scores across all datasets, demonstrating \model's superior generalizability across diverse geographic regions. For instance, on GeoLife, \model's fine-tuned performance (MAE 23.23) reduces error by 32.73\% compared to TrajBERT, showcasing its effectiveness with complex travel patterns and lower-quality data.
These results validate \datasetname's potential as a foundation dataset and \model's consistent superiority in trajectory recovery tasks, with substantial improvements through fine-tuning, reinforcing its adaptability and robustness.

\begin{wraptable}{r}{7.5cm}
    \vspace{-18pt}
    \caption{\fontsize{9.5pt}{12pt}\selectfont{Performance comparison of \model with trajectory prediction tasks.}}
      \fontsize{9pt}{12pt}\selectfont
      \resizebox{0.53\textwidth}{!}{
\begin{tabular}{lcc cc  cc} 
\toprule
\multirow{2}{*}[\multirowoffset]{\textbf{Methods}}  & \multicolumn{2}{c}{\textbf{WorldTrace} } & \multicolumn{2}{c}{\textbf{Chengdu}} & \multicolumn{2}{c}{\textbf{GeoLife}}\\
\cmidrule(lr){2-3}\cmidrule(lr){4-5}\cmidrule(lr){6-7}
    & \textbf{MAE}  & \textbf{RMSE}  & \textbf{MAE}  & \textbf{RMSE}  & \textbf{MAE}  & \textbf{RMSE}   \\ 
\cmidrule(lr){1-7} 
Linear  & 153.12 & 159.65 & 156.85 & 164.58 & 189.02 & 201.34 \\
DHTR  & 146.48 & 151.63	& 123.47 & 129.73	& 180.32 & 187.59 \\
Transformer & 114.25 & 117.07 & 67.38 & 70.86 & 165.02 & 170.84\\
DeepMove  & 55.69 & 58.67 & \underline{36.31} & \underline{39.10} & 116.46 & 123.20  \\
TrajBERT   & 80.57 & 86.36  & 64.73	 & 68.92 & 113.68  & \underline{121.18}    \\
TrajFM  & 75.45 & 81.32 & 77.82 & 80.48 & 121.94 & 128.16 \\
\cmidrule(lr){1-7}
\model (zero-shot) & \underline{49.85} & \underline{55.02} & 42.75 & 45.93 & \underline{108.35} & 133.60 \\
Improvement(\%) & \red{10.49} & \red{6.22} & \blue{17.74} & \blue{17.46} & \red{4.69} & \blue{10.25} \\
\model (fine-tune) & \textbf{30.10} & \textbf{34.46} & \textbf{28.78} & \textbf{32.44}	& \textbf{90.97} & \textbf{102.88} \\
Improvement(\%) & \red{45.95} & \red{41.27} & \red{20.74} & \red{17.03} & \red{19.98} & \red{15.10} \\
\bottomrule
\end{tabular}
\label{tab:prediction}
}
\vspace{-5mm}
\end{wraptable}
\textbf{Trajectory Prediction}.
Table~\ref{tab:prediction} shows UniTraj's exceptional performance in trajectory prediction, a different task requiring forward inference rather than reconstruction. 
The zero-shot results merit particular attention, as they represent the most challenging scenario for trajectory models.
On WorldTrace, UniTraj's zero-shot MAE significantly outperforms all baselines, underscoring the model's versatility in capturing universal motion patterns.
When fine-tuned, the performance further improves, consistently achieving the best results across all evaluated datasets.
This generalization capability stems from our Last-N masking strategy, which explicitly shapes the embedding space to support predictive inference. 
These results further confirm that \model not only generalizes remarkably well across diverse datasets but also benefits considerably from fine-tuning, making it highly adaptable for real-world applications requiring accurate trajectory predictions.

\begin{wrapfigure}{r}{0.5\textwidth}
\vspace{-5mm}
    \centering
    \includegraphics[width=0.5\textwidth]{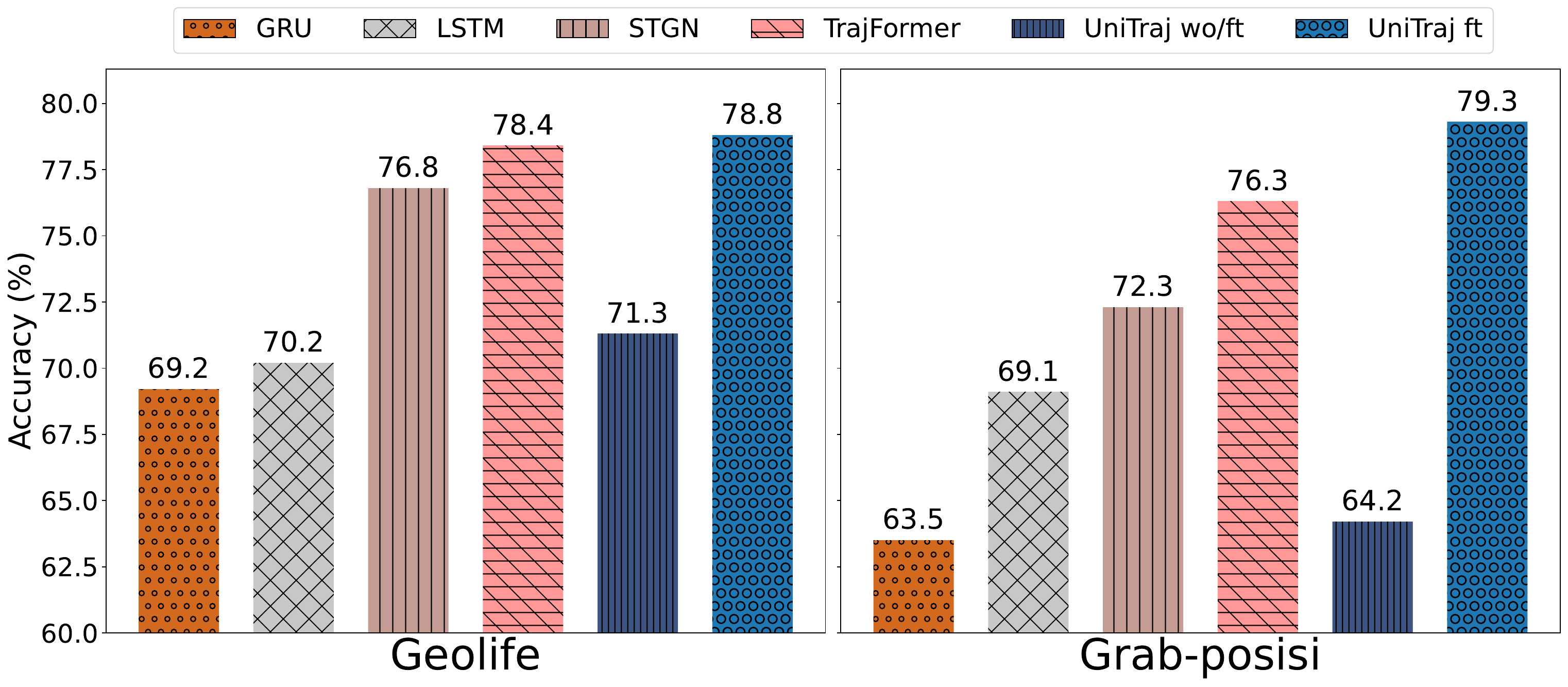}
    \vspace{-5mm}
    \caption{Performance comparison of classification task with GeoLife and Grab-posisi dataset.}
    \vspace{-3mm}
    \label{fig:classification}
\end{wrapfigure}

\textbf{Trajectory Classification}.
Figure~\ref{fig:classification} presents classification accuracy results that reveal UniTraj's capacity to learn discriminative representations of movement modalities.
Notably, even without fine-tuning, UniTraj achieves 71.3\% accuracy on GeoLife, outperforming several supervised baselines. This zero-shot performance demonstrates that the pre-trained representations inherently capture transportation mode signatures, where movement modality emerges as a natural organizing principle. 
On the Grab-Posisi dataset, which presents additional challenges due to similar motion patterns for mixed travel modes (car and motorcycle).
UniTraj achieves 79.3\% accuracy after fine-tuning with a substantial improvement over the best baseline. 
This improvement emphasizes UniTraj's ability to capture subtle kinematic signatures that differentiate travel modes with complex or similar patterns.

\subsection{Dataset Study}
This section analyzes the impact of dataset scale, quality, and diversity on model performance of UniTraj, particularly its generalization capability across different data sources. We focus on two main experiments: (1) examining the effect of dataset scale and quality within \datasetname, with varying data volumes ($\{0.01,~0.5,~1\}$ millions) and a high-quality  (obtained by further removing loops, staying dense trajectories) subset, and (2) assessing UniTraj’s adaptability and effectiveness by training it on these datasets beyond \datasetname, thus showing its potential as a foundation model.

\begin{figure}[h]
    \centering
     \subfigure[Training data volume.]{
        \includegraphics[width=0.23\linewidth]{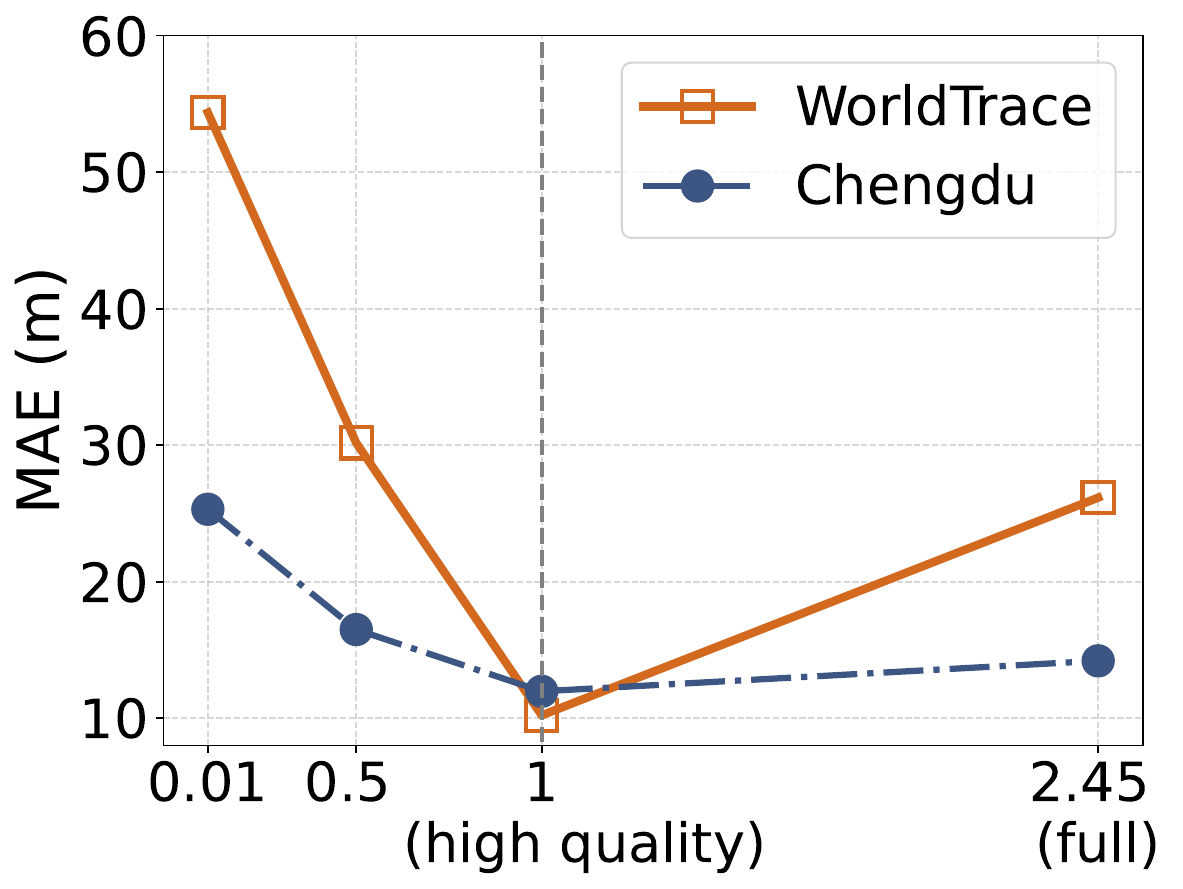}
        \label{fig:datavolume_study}
     }
        \subfigure[Training dataset.]{
        \includegraphics[width=0.23\linewidth]{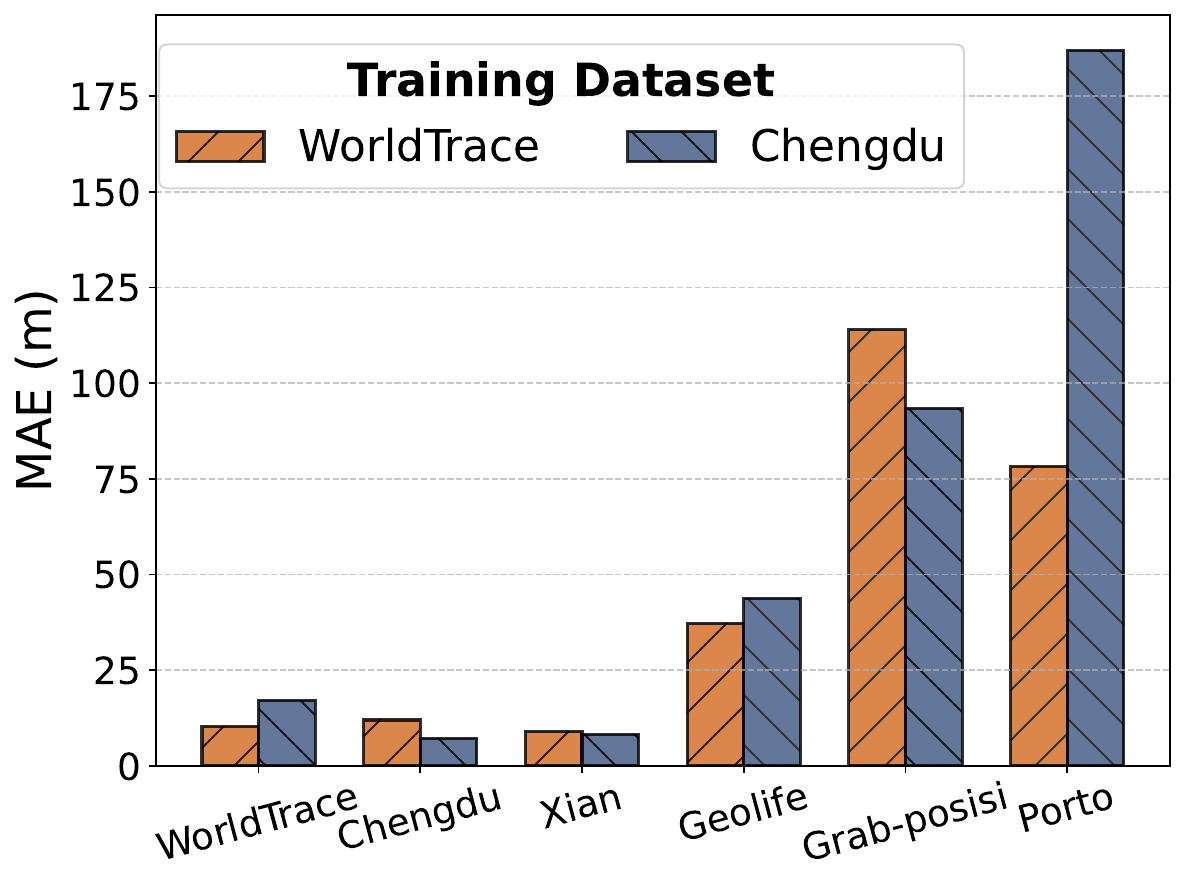}
        \label{fig:dataset_study}
    }
    \subfigure[Number of Encoders]{
        \includegraphics[width=0.23\linewidth]{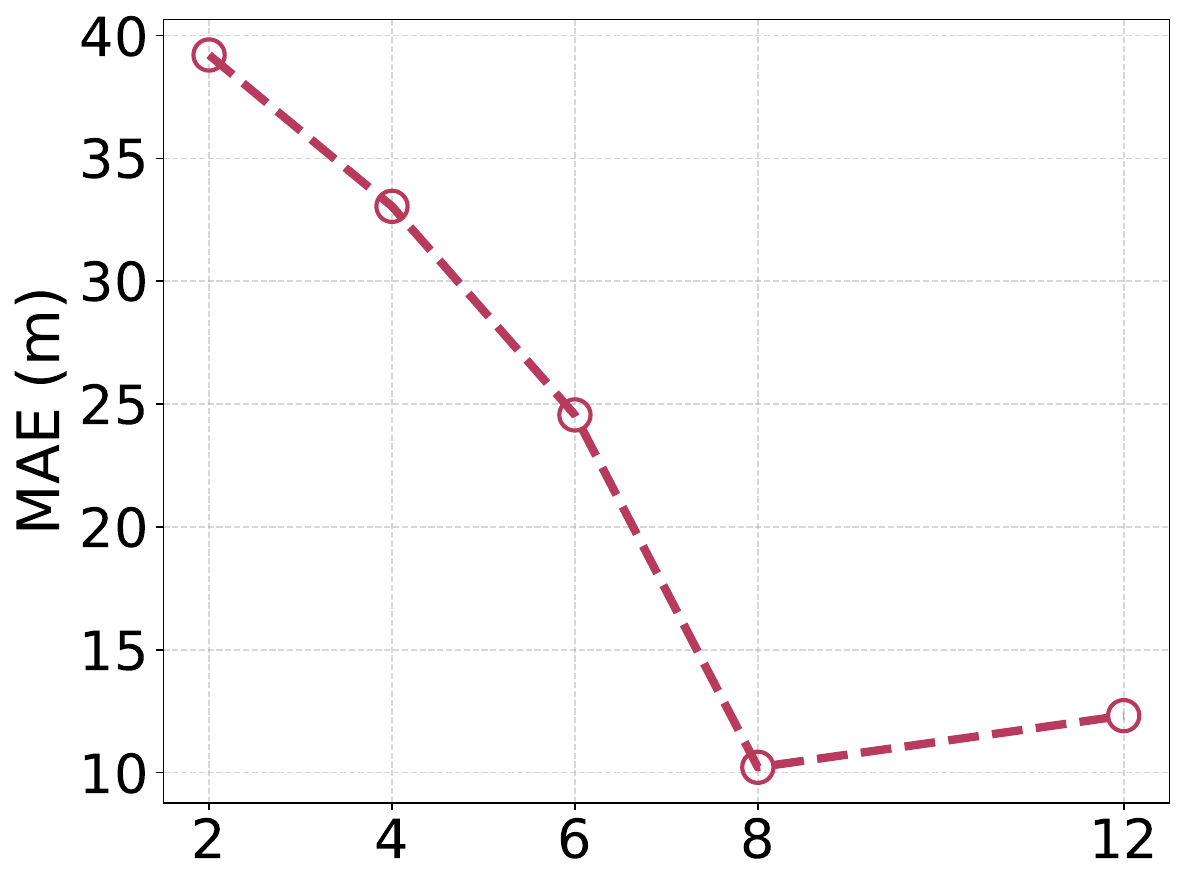}
        \label{fig:encoders_sensitivity}
     }
     \subfigure[Mask ratio (\%).]{
        \includegraphics[width=0.23\linewidth]{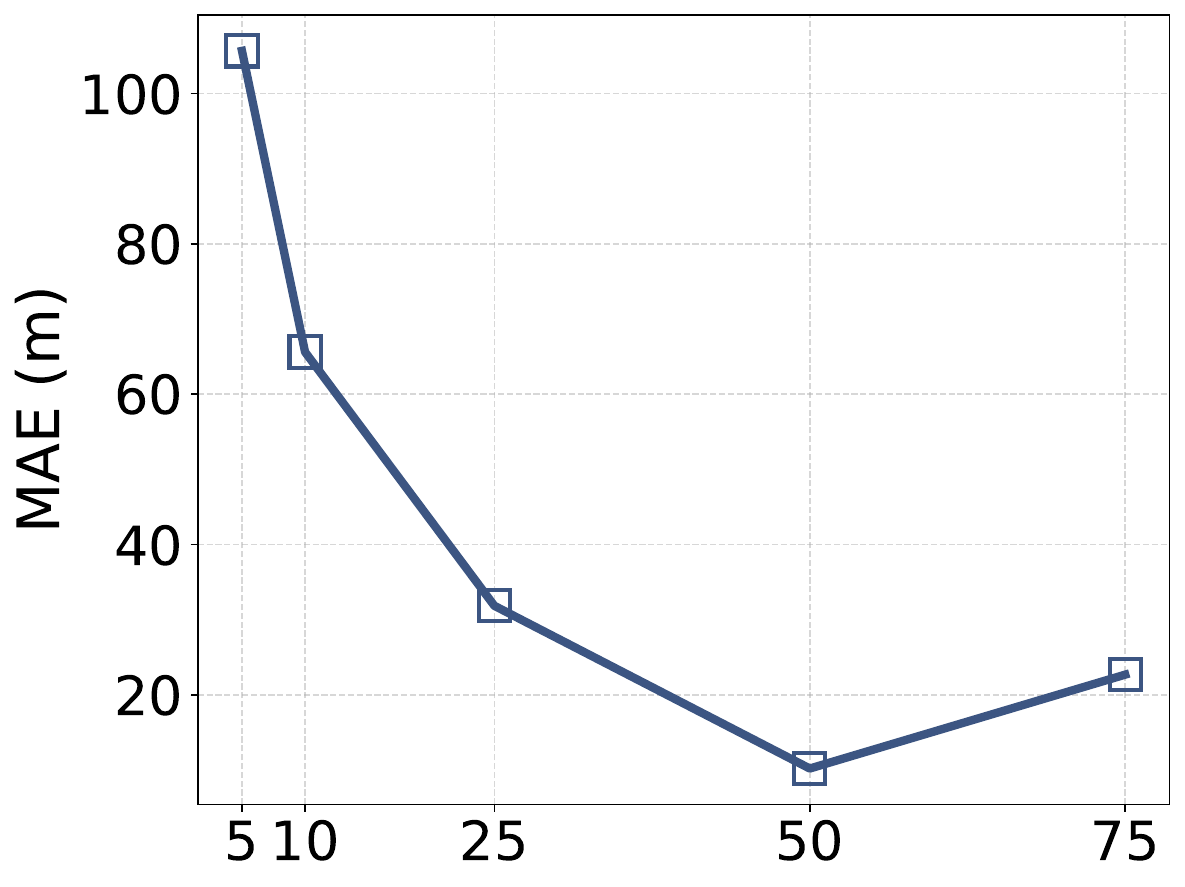}
        \label{fig:ratio_sensitivity}
    }
    \vspace{-3mm}
    \caption{The effect of amount of data volume, diversity dataset, and different parameter settings.
    }
   \label{fig:para_study}
\end{figure}

{\textbf{Effect of Dataset Scale and Quality}.} 
Figure~\ref{fig:datavolume_study}  illustrates the relationship between training data volume and model performance, revealing a phenomenon that goes beyond simple scaling laws. 
With increasing trajectory count from WorldTrace (from 0.5M to 2.45M), the MAE on the in-domain test set decreases dramatically, showing substantial improvement up to approximately 1M trajectories before beginning to exhibit diminishing returns. 
The above result indicates that larger datasets enable the model to capture a wider range of spatio-temporal patterns. However, while increasing the dataset size from 1 million to 2.45 million trajectories results in better coverage, the model’s MAE slightly increases due to the introduction of more noise in the full dataset. In contrast, training on a high-quality subset of 1 million trajectories, which includes curated, noise-free data, yields more reliable and consistent learning. This highlights the importance of both dataset scale and quality, with quality being especially crucial when data volume is limited.

{\textbf{Effect of Dataset Diversity}.} 
In Figure~\ref{fig:dataset_study}, we compare \model’s zero-shot performance when trained on \datasetname and Chengdu (the highest quality dataset available), evaluated across multiple real-world datasets. Models trained on \datasetname exhibit superior generalization across diverse datasets (e.g., GeoLife and Porto), reflecting the broad geographic and contextual coverage of \datasetname. Conversely, models trained on Chengdu perform best on datasets with similar density and travel modes, such as Xi’an. However, proprietary datasets like Chengdu, while offering high quality, are not publicly available, limiting their applicability for universal tasks. These results demonstrate UniTraj’s robustness and adaptability, validating \datasetname as an ideal training resource for building a universal trajectory foundation model. At the same time, the findings confirm that \model can effectively leverage other datasets when necessary, further enhancing its versatility.

\subsection{Model Study}\label{sec:modelstudy}
We investigate architectural components, parameter settings, and pre-training strategies to assess sensitivity to parameter choices and the contributions of their core components.

{\textbf{Effect of Parameter Settings}.} 
Figure~\ref{fig:encoders_sensitivity} Figure~\ref{fig:ratio_sensitivity}  and presents the results of our parameter sensitivity analysis, examining how the number of encoder blocks and the masking ratio influence model performance. As shown in Figure~\ref{fig:encoders_sensitivity}, increasing the number of encoder blocks from 2 to 8 significantly reduces MAE, with performance plateauing beyond 8 blocks. This plateau suggests that while deeper architectures can improve model capacity, the benefits diminish without corresponding adjustments in data or hyperparameters~\cite{kaplan2020scaling}.
Figure~\ref{fig:ratio_sensitivity} demonstrates that a masking ratio of 50\% yields the best performance. Low masking ratios (e.g., 5\%-10\%) result in higher MAE due to insufficient training signal, while higher ratios (e.g., 75\%) lead to increased MAE from excessive information loss. A 50\% masking ratio strikes a balance, providing the model with a strong training signal without sacrificing the context needed for effective trajectory reconstruction.

\begin{table*}[t]
\small
\caption{Ablation study on different resampling and masking strategies on six datasets.}
\centering
\resizebox{0.99\textwidth}{!}{
\begin{tabular}{lcc cc  cc cc cc cc} 
\toprule
\multirow{2}{*}[\multirowoffset]{\textbf{Methods}}  & \multicolumn{2}{c}{\textbf{WorldTrace} } & \multicolumn{2}{c}{\textbf{Chengdu}} & \multicolumn{2}{c}{\textbf{Xi'an}}  & \multicolumn{2}{c}{\textbf{GeoLife}} & \multicolumn{2}{c}{\textbf{Grab-posisi}} & \multicolumn{2}{c}{\textbf{Porto}} \\
\cmidrule(lr){2-3}\cmidrule(lr){4-5}\cmidrule(lr){6-7}\cmidrule(lr){8-9}\cmidrule(lr){10-11}\cmidrule(lr){12-13}
     & \textbf{MAE}  & \textbf{RMSE}  & \textbf{MAE}  & \textbf{RMSE}  & \textbf{MAE}  & \textbf{RMSE} & \textbf{MAE}  & \textbf{RMSE}  &  \textbf{MAE} & \textbf{RMSE} &  \textbf{MAE} & \textbf{RMSE}\\ 
\cmidrule(lr){1-13} 
w/o Dynamic Multi-scale resampl. & 426.80 & 482.37 & 192.54 & 272.42 & 157.85 & 223.96 & 499.95 & 671.69 & 1933.28 & 2504.16 & 93.14 & 119.93 \\
w/o Interval Consistent resampl. &  21.30 & 24.76 & 12.98 & \underline{20.61} & 9.34 & 13.90 & 69.41 & 115.33 & \underline{102.45} & \underline{149.60} & 1724.12 & 2016.61  \\
w/o Key points masking & 25.49 & 28.91 & 14.46 & 21.98 & 11.10 & 15.17 & \underline{45.94} & \underline{72.84} & 113.65 & 162.57 & \textbf{76.51} & \underline{101.18} \\
w/o Block masking & \textbf{7.79} & \textbf{10.47} & \textbf{9.22} & \textbf{15.36} & \textbf{7.16} & \textbf{11.18} & 48.59 & 77.73 & \textbf{89.34} & \textbf{128.72} & 198.41 & 238.88 \\
\model & \underline{10.22} & \underline{13.56} & \underline{11.98} & 20.94  & \underline{8.93} & \underline{13.83} & \textbf{37.21} & \textbf{63.89} & 114.07 & 167.01 & \underline{78.28} & \textbf{100.14} \\
\bottomrule
\end{tabular}
}
\vspace{-3mm}
\label{tab:ablation}
\end{table*}

{\textbf{Ablation Study}.} 
Table~\ref{tab:ablation} presents an ablation study, showing how different pre-training strategies affect UniTraj's performance across datasets. The performance varies across datasets, indicating the effectiveness and limitations of them depending on the specific data and task scenarios.
\emph{Dynamic Multi-scale Resampling} significantly improves performance across most datasets, especially GeoLife and Grab-Posisi, which have inconsistent sampling intervals and lower data quality. 
This suggests that dynamic resampling helps the model to adapt to heterogeneous dataset scenarios and to be adaptive for information preservation (see \textbf{Appendix \ref{app:DMSR}} for more details).
The \emph{Interval Consistent Resampling} has a notable positive effect on datasets with consistent sampling rates, such as Porto and \datasetname. 
It indicates that the integration of this strategy strategy can effectively separate the temporal sampling pattern from the region, it enhances the generalization of the model to data sets with different sampling rates (analysis presented in \textbf{Appendix \ref{app:ICR}}).
\emph{Key Points Masking} leads to substantial performance drops on high-quality datasets like Chengdu and Xi’an but appears to offer minimal benefits, or even slight disadvantages, for certain datasets. 
This finding suggests that adjusting adaptive masking strategies based on trajectory complexity, potentially applying it selectively to trajectories with significant directional changes, while using alternative strategies for smoother paths.
\emph{Block Masking} shows significant effects on GeoLife and Porto, where it helps the model handle low sampling frequencies. However, its impact on other datasets is more inconsistent, suggesting that it introduces an artificial challenge that may increase complexity in high-frequency datasets. (we provide a robustness analysis in \textbf{Appendix \ref{app:masking}})
Overall, the varying impact of UniTraj’s pre-training strategies across datasets highlights its adaptability to different tasks and scenarios. While not all of them universally enhance performance, their combined use provides a balanced training strategy, allowing for flexible configuration depending on specific dataset requirements. Fine-tuning further optimizes performance, ensuring stability and robustness across diverse tasks.

\section{Conclusion}\label{sec:conclusion}

In this work, we presented \model, a universal trajectory foundation model designed to overcome the task specificity, regional dependency, and data quality limitations of current approaches. \model acts as a robust backbone that generalizes effectively across diverse tasks and regions.
To support its development, we introduced \datasetname, a high-quality global dataset with 2.45 million trajectories from 70 countries, offering broad geographic coverage, varied sampling rates, and open accessibility. 
Together, \model and \datasetname provide a versatile, high-performing foundation for trajectory analysis, paving the new solution for more adaptable and efficient models in trajectory-based research. 
Future work will focus on expanding the geographic and modal diversity of the WorldTrace dataset to better cover underrepresented regions and non-motorized travel. We also aim to enhance the UniTraj model by integrating contextual information, such as road networks and points of interest, to improve its predictive accuracy and real-world applicability. Further optimizations to the model architecture and pre-training strategies will also be explored to boost performance and efficiency.

\bibliography{ref}
\bibliographystyle{abbrv}

\clearpage
\addtocontents{toc}{\protect\setcounter{tocdepth}{2}}

\appendix
\onecolumn

{\centering
  \Large{\Huge\scshape Supplementary Material\par}
  {\large \sc UniTraj: Learning a Universal Trajectory Foundation Model \\  from Billion-Scale Worldwide Traces\par}
}
\vskip 4mm
\startcontents[sections]\vbox{\sc\Large Table of Contents}
\vspace{5mm}
\hrule height .8pt
\vspace{-2mm}
\printcontents[sections]{l}{1}{\setcounter{tocdepth}{2}}
\vspace{4mm}
\hrule height .8pt
\vskip 10mm

\clearpage
\section{Details of WorldTrace Dataset}\label{app:dataset}
In this section, we detail the collection of the dataset, the processing, and provide a detailed analysis of the resulting dataset.

\begin{figure}[h]
\centering
    \includegraphics[width=0.9\linewidth]{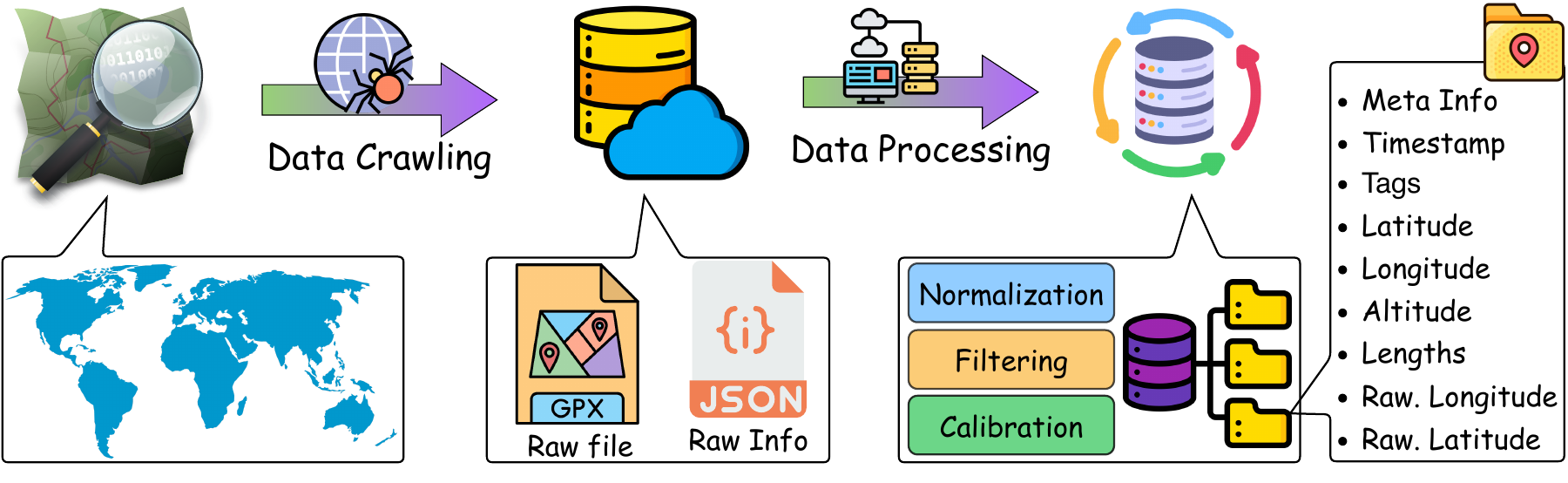}
    \caption{The process pipeline of WorldTrace dataset construction.}
    \label{app_fig:pipeline}
\end{figure}

\subsection{Data Collection}\label{sec:collection}
\textbf{Data Source.} As shown in Figure \ref{app_fig:pipeline}, the raw data for WorldTrace is sourced from the shared trajectory data platform on OpenStreetMap (OSM) \cite{OpenStreetMap}\footnote{https://www.openstreetmap.org/traces}. This platform, a public sharing project, hosts over 11 million GPS trajectories uploaded by contributors worldwide from 2004 to the present.
To ensure data quality and reliability, we specifically targeted contributions tagged for motorized movement to ensure data currency and relevance to modern transportation networks. 
This approach helps minimize device heterogeneity and avoids outdated data that might not reflect current infrastructure.
The raw data is stored in the standardized GPX (GPS Exchange Format), an XML schema designed for exchanging GPS data between applications and web services \href{https://www.topografix.com/GPX/1/1/}{https://www.topografix.com/GPX/1/1/}. Each GPX file contains sequences of trackpoints with the following attributes:
\begin{itemize}[leftmargin=*]
    \item Latitude (decimal degrees)
    \item Longitude (decimal degrees)
    \item Altitude (decimal numbers)
    \item Timestamp (ISO 8601 format)
    \item Optional metadata (version, tags, etc.)
\end{itemize}
In addition, while crawling the original trajectory, we also crawled the basic information about the trajectory descriptions, such as the starting point, markers, time, creator, etc., which was saved as a JSON file.

\textbf{Collection Process}.
Prior to integration, our collection pipeline involved the following steps:
Our collection pipeline involved the following steps:
\begin{enumerate}[leftmargin=*]
    \item \textbf{API-based Retrieval}: We use the OSM API to systematically query and download GPX traces based on selected filters to ensure global coverage. In order not to increase the burden on server providers, we did not use concurrent crawling, and the whole collection process lasted about 6 months, yielding about 4.5 million raw traces.

    \item \textbf{Initial Filtering}: During acquisition, we implemented preliminary filtering to exclude trajectories with obvious anomalies such as: Coordinates outside valid ranges (-90° to 90° for latitude, -180° to 180° for longitude);
    Duplicate or long duration consecutive points; Empty or near-empty traces (fewer than 60 seconds).

    \item \textbf{Format Standardization}: All collected data was parsed from the original GPX format and converted to a unified internal format for subsequent processing.
\end{enumerate}

\subsection{Data Processing} 
Our preprocessing pipeline was designed to balance preserving authentic movement patterns with removing noise and inconsistencies. The process consists of three main stages:

\textbf{Normalization}.
The raw data exhibited highly variable sampling frequencies, ranging from sub-second intervals up to several seconds between consecutive points. This heterogeneity creates challenges for modeling and increases storage requirements unnecessarily. We therefore applied the following normalization procedures:
\begin{itemize}[leftmargin=*]
    \item \textbf{Temporal Resampling}: We resampled all trajectories to a uniform rate of one point per second (1 Hz). For segments with sampling rates higher than 1 Hz, we select the first occurrence of a trajectory point within each one-second window. For segments with lower sampling rates, we used linear interpolation between available points to estimate positions at one-second intervals.

    \item \textbf{Coordinate Standardization}: All coordinates were converted to the WGS84 datum for consistency, and we ensured uniform precision across the dataset (6 decimal places for both latitude and longitude, providing ~0.1m precision at the equator).
\end{itemize}

\textbf{Filtering}.
After normalization, we implemented a multi-stage filtering process to meticulously remove trajectories that were deemed unsuitable for our analysis. This comprehensive filtering approach involved several key steps:
\begin{itemize}[leftmargin=*]
    \item \textbf{Length-based Filtering}: We discarded trajectories with fewer than 32 points (equivalent to 32 seconds after resampling) or covering distances below 100 meters, as these typically represent stationary periods or very short movements with limited analytical value.
    \item \textbf{Speed-based Filtering}: We calculated point-to-point speeds and removed trajectories containing implausible values (e.g., exceeding 120 km/h or lower 0.5 km/h in urban environments), typically caused by GPS errors or anomalies.

    \item  \textbf{Distance-based Outlier Detection}:We calculated the distance between the original trajectory and the map-matched trajectory. Trajectories that were too far away (indicating large deviations in motion) were flagged for further inspection or removal.
    \item \textbf{Loop Detection}: We identify and remove trajectories that form perfect or near-perfect loops with no apparent destination by their geometry, which usually indicates the presence of clearly anomalous patterns.
    
\end{itemize}

\textbf{Calibration}.
GPS signals can suffer from various errors due to atmospheric conditions, satellite geometry, and physical obstructions. To improve data quality, we applied map-matching techniques to align raw GPS points with underlying road networks, using a Hidden Markov Model-based approach (or using online API) with a custom emission probability function that accounts for both point-to-road distance and heading consistency.
Besides, each trajectory point was enriched with derived attributes.

\begin{figure}[t]
\centering
    \subfigure[Geographic distribution across the world.]{
    \includegraphics[width=0.92\linewidth]{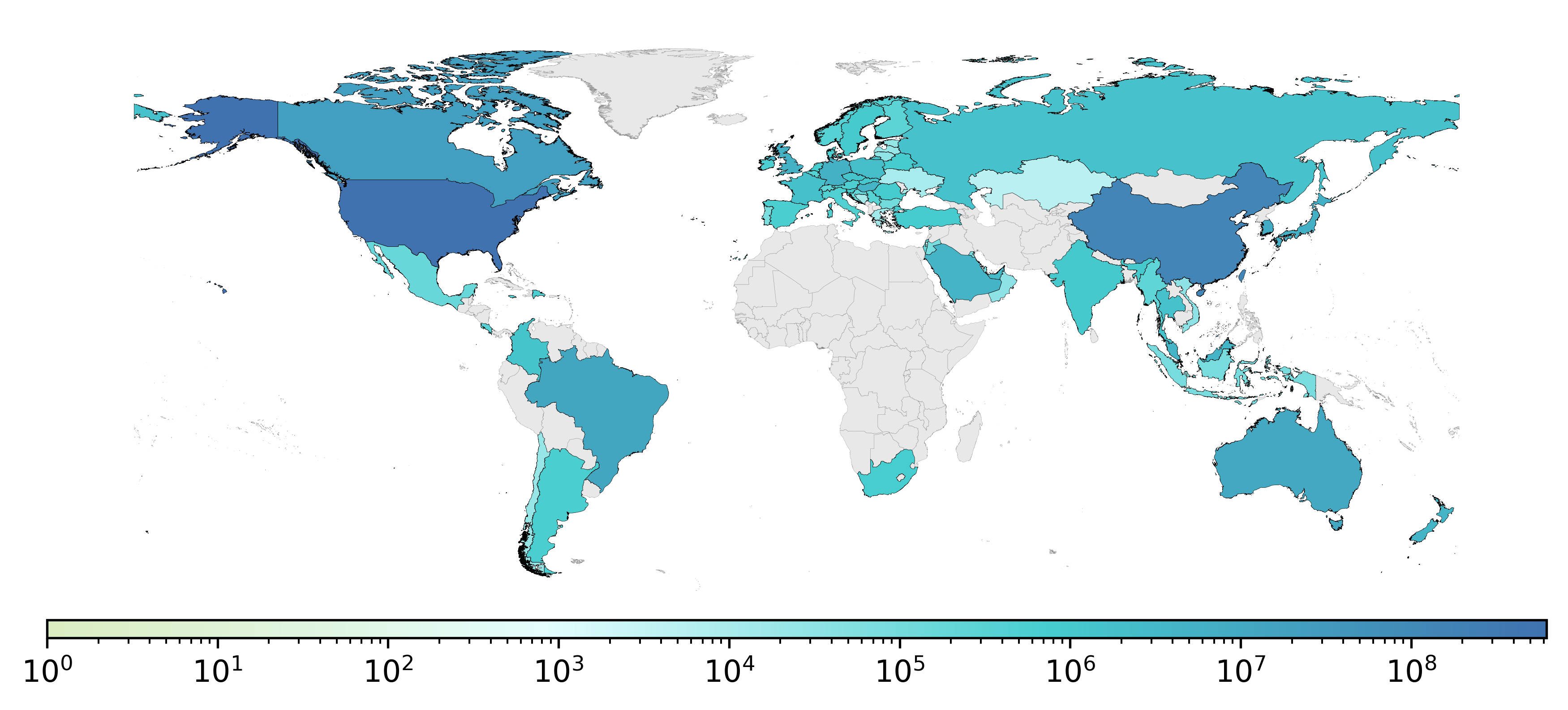}
    \label{fig:geo_dis}
    }
    \subfigure[Trajectory counts of top 10 countries.]{
        \includegraphics[width=0.45   \linewidth]{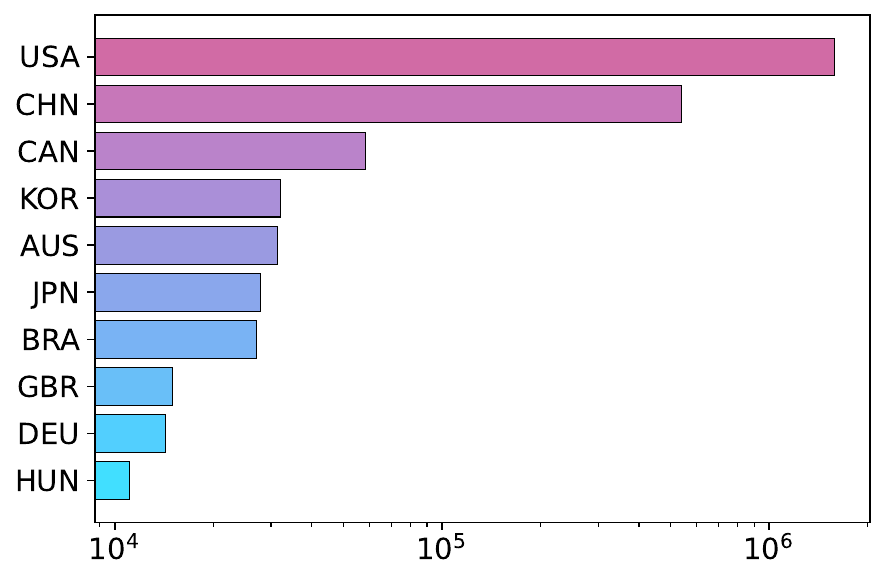}
        \label{fig:geo_top10}
    }
        \subfigure[Distribution within contiguous USA.]{
        \includegraphics[width=0.45\linewidth]{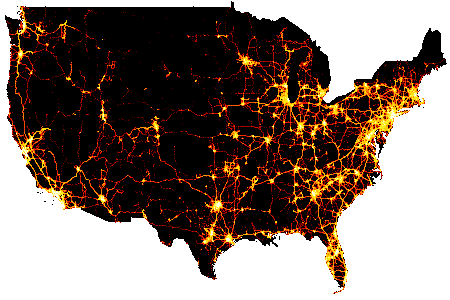}
        \label{fig:geo_us}
    }
   \caption{The distribution details of WorldTrace dataset.}
       \label{fig:WorldTrace}
\end{figure}

\subsection{Data Statistics and Analysis}
\textbf{Overall Statistics}.
The final WorldTrace dataset contains:
\begin{itemize}[leftmargin=*]
    \item Approximately 2.45 million trajectories.
    \item 8.8 billion raw GPS points (before normalization).
    \item Coverage across 70 countries on all inhabited continents.
    \item Temporal span from August 2021 to December 2023.
    \item Average trajectory duration of approximately 6 minutes.
    \item Average trajectory distance of 5.73 kilometers.
    \item Average travel speed of 48.0 km/h.
    \item Points per trajectory ranging from 32 to over 600, with an average of 358 points.
\end{itemize}

\textbf{Geographic Distribution}.
WorldTrace offers extensive geographic coverage, as illustrated in Figure~\ref{fig:WorldTrace}, encompassing trajectory data from 70 countries and spanning diverse environments and infrastructure types. This global distribution is visualized in Figure~\ref{fig:geo_dis}, highlighting dense concentrations in North America, East Asia, and parts of Europe, with trajectory counts exceeding  in the most represented regions.
Figure~\ref{fig:geo_top10} further details the top 10 countries by trajectory counts, with the United States, China, and Canada leading in data volume. 
Notably, it exhibits substantial geographic diversity, with varying densities across urban, suburban, and rural environments. 
The top 10 countries by trajectory count, namely, the USA, China, Canada, Germany, UK, Japan, Brazil, Australia, South Korea, and Hungary, represent a wide range of urban forms, road networks, and mobility cultures.
Additionally, Figure~\ref{fig:geo_us} provides a closer look at the data density within the contiguous United States, demonstrating high-resolution coverage along major road networks and urban centers. 
This detailed distribution underscores the dataset's ability to capture nuanced variations in trajectory data across different regions.
Collectively, these figures emphasize the potential of WorldTrace to serve as a robust foundation for developing region-independent and universal trajectory models. Its extensive geographic coverage and diverse environmental representation make it well-suited for applications that require broad and adaptable trajectory data.

\subsection{Data Privacy and Copyright}
To protect privacy and comply with international data protection regulations, all data collection adhered strictly to privacy regulations and ethical guidelines. Trajectories were anonymized, and any personally identifiable information was excluded to protect user privacy. In addition, all raw data follows the Open Data Commons Open Database License (\textbf{ODbL}) license
from OSM: \href{http://opendatacommons.org/licenses/odbl/1.0/}{http://opendatacommons.org/licenses/odbl/1.0/}. 
We will share derived datasets under the same license terms to respect the
data use policies of the community.

\section{Pre-training Strategies}\label{app:pretraining}

In this section, we provide specific details on the adaptive trajectory resampling strategy and the self-supervised trajectory masking strategy, and we will provide the design motivation and theoretical analysis for these two strategies.

\subsection{Adaptive Trajectory Resampling}\label{app:resampling}

Trajectory data heterogeneous is one of the main challenges in cross-regional and cross-device trajectory modeling. The Adaptive Trajectory Resampling strategy solves this problem through two complementary components: Dynamic Multi-Scale Resampling and Interval Consistent Resampling.
We designed these two strategies with the motivation of fitting different regions and dataset qualities through diversified trajectory sampling frequencies and motion patterns.
\textit{Dynamic Multi-Scale Resampling ensures an optimal balance between information preservation and computational efficiency across different trajectory lengths, prioritizing the retention of key motion patterns.
Interval Consistent Resampling enhances the model's generalization ability across datasets with different sampling rates by normalizing the time dimension.}

\subsubsection{Dynamic Multi-Scale Resampling}\label{app:DMSR}
As discussed in Section \ref{sec:pre-training}, we adopted a logarithmic resampling ratio that adjusts the sampling rate according to the trajectory length. 
The resampling ratio function $R(n)$ is designed to decrease logarithmically as the trajectory length $n$ increases:
\begin{align}\label{eq:app_sampling}
R(n) =
\begin{cases}
R_{\text{min}}, & n \geq n_{\text{max}} \\
1- (1 - R_{\text{min}})  \phi(n), & n_{\text{min}} < n < n_{\text{max}} \\
1, &  n \leq n_{\text{min}}
\end{cases}
\end{align}
where $R_{\text{min}}$ is the minimum sampling ratio, and $n_{\text{min}}$ and $n_{\text{max}}$ denotes the shortest and longest length thresholds, respectively.
The normalization factor $\phi(n)$ is computed as follows:
\begin{align}\label{eq:app_phi}
\phi(n) = \frac{\ln\left( n - n_{\text{min}} + 1 \right)}{\ln\left( n_{\text{max}} - n_{\text{min}} + 1 \right)}.
\end{align}

\textbf{Formal Definition}.
Here, we provide a formal definition and theoretical analysis of the above empirical results through information theory and computational efficiency perspectives.
For any trajectory  $\boldsymbol{\tau}=\{p_1,~p_2,~\ldots,~p_n\}$ consist of $n$ spatio-temporal points. the number of points for resampled trajectory $\boldsymbol{\tau}^{\prime} = \{ p_1, p_2, \dots, p_m \}$ is:
\begin{align}
m = R(n) \cdot n,
\end{align}
where function $R(n)$ that determines what proportion of points to retain.
The logarithmic sampling strategy guarantees bounded sample sizes for arbitrarily long trajectories while preserving critical minimum information content.
Specifically:
\begin{itemize}[leftmargin=*]
    \item For $n \leq n_{\text{min}}$: $R(n) = 1$, so $m = n$;
    \item For $n \geq n_{\text{max}}$: $R(n) = R_{\min}$, we set $m = m_{\max}$ as a constant. Clearly, the number of sampled points is bounded above by $m_{\max}$.
\end{itemize}
To ensure boundedness, we analyze $m(n)$ in the intermediate domain $n \in (n_{\text{min}}, n_{\text{max}})$. 
\begin{align}
m = \bigl[1 - (1 - R_{\min}) \cdot \phi(n) \bigr] \cdot n.
\end{align}
Taking derivative:
\begin{align}
\frac{d(R(n) \cdot n)}{dn} = 0
\end{align}
Solving this equation yields a value $n^* < n_{\max}$, ensuring that $m_{\max}$ is bounded.
Since $R(n)$ becomes constant for $n \geq n_{\max}$, and $m$ increases linearly in that region, the global maximum occurs at either $n^*$ or $n_{\max}$.
However, due to the logarithmic decay of $R(n)$, the growth of $m$ slows, and the maximum value is achieved at a finite $n^* < n_{\max}$. Hence, $m$ is bounded for all $n$.




{\large{\textit{\textbf{Corollary 1}: Information Preservation and Computing Efficiency Optimization}}}

\textbf{Standpoint:} The logarithmic sampling function provides an optimal balance between information preservation and computational efficiency across varying trajectory lengths.

\textit{Proof:}
Let $I(\boldsymbol{\tau})$ represent the information content of trajectory $\boldsymbol{\tau}$. Empirical studies in spatio-temporal data analysis suggest that information content typically scales sub-linearly with trajectory length, following approximately:
\begin{align}\label{eq:info_cont}
I(\boldsymbol{\tau}) \propto n^{\alpha},
\end{align}
where $0 < \alpha <1$. For example, $\alpha \approx 0.7$ indicates that only 70\% of the trajectory points contain valid feature information, and the remaining 30\% are redundant.
For a resampled trajectory $\boldsymbol{\tau}^{\prime}$ with $m = R(n) \cdot n$ points, the information preservation ratio $\eta$ can be approximated as:
\begin{align}
    \eta = \frac{I(\boldsymbol{\tau^{\prime}})}{I(\boldsymbol{\tau})} \approx (\frac{m}{n})^{\alpha} = R(n)^{\alpha}.
\end{align}
The computational cost $C$ of processing trajectory typically scales linearly with length:
\begin{align}
      C(\boldsymbol{\tau}) \propto n^{\beta},
\end{align}
where $\beta \ge 1$, typically  $\beta \approx 2$ for Transformer-based models.
After resampling, the computational efficiency gain $\gamma$ is:
\begin{align}
    \gamma = \frac{C(\boldsymbol{\tau})}{C(\boldsymbol{\tau^{\prime}})} \approx (\frac{n}{m})^{\beta} = \frac{1}{R(n)^{\beta}} .
\end{align}
The optimal sampling function maximizes the product of information preservation and computational efficiency:
\begin{align}
    \max_{R(n)} \eta \cdot \gamma = \max_{R(n)} R(n)^{\alpha} \cdot \frac{1}{R(n)^{\beta}} = \max_{R(n)}  R(n)^{\alpha - \beta}.
\end{align}
Since $\alpha < \beta$ for typical trajectory data, this is a decreasing function $R(n)$. However, we must maintain a minimum level of information, hence the constraint $R(n) \ge R_{\min}$.

When we examine the information density:
\begin{align}
    D(n) = \frac{I(\boldsymbol{\tau})}{n} \propto n^{\alpha -1},
\end{align}
we observe that it decreases as $n$ increases, indicating diminishing information return per point in longer trajectories. 
An optimal sampling ratio should proportionally track this information density: 
\begin{align}
R_{\text{opt}}(n) \propto D(n) \propto n^{\alpha -1}.
\end{align}
Our logarithmic resampling function's derivative in the intermediate domain ($n_{\min}<n < n_{\max}$) is: 
\begin{align} 
\frac{dR(n)}{dn} = -\frac{1-R_{\min}}{\ln(n_{\max}-n_{\min}+1)} \cdot \frac{1}{n-n_{\min}+1} \propto \frac{1}{n} 
\end{align}
As $n$ increases, the growth rate of the logarithmic function slows down, causing the rate at which the sampling rate $R(n)$ decreases to also slow down. 
This is closely proportional to the derivative of the theoretical optimal sampling rate:
\begin{align}
    \frac{d R_{\text{opt}}(n)}{dn} \propto (\alpha -1) n^{\alpha-2} \propto\frac{1}{n^{2-\alpha}}.
\end{align}
For example, when $\alpha \approx 0.7$, we have $\frac{d R_{\text{opt}}(n)}{dn}  \propto \frac{1}{n^{1.3}}$
This property of logarithmic functions (their rate of change is inversely proportional to the input value), making them naturally suited to this task. 
Therefore, Logarithmic resampling provides a theoretically reasonable compromise: it preserves almost all of the information from short trajectories (where every point may be significant) while reducing redundancy in long trajectories (where redundancy is highest). 
Compared to linear functions, logarithmic functions can more naturally adapt to the information density curve across the entire trajectory length range.




\begin{figure}[h]
\centering
    \includegraphics[width=0.7\linewidth]{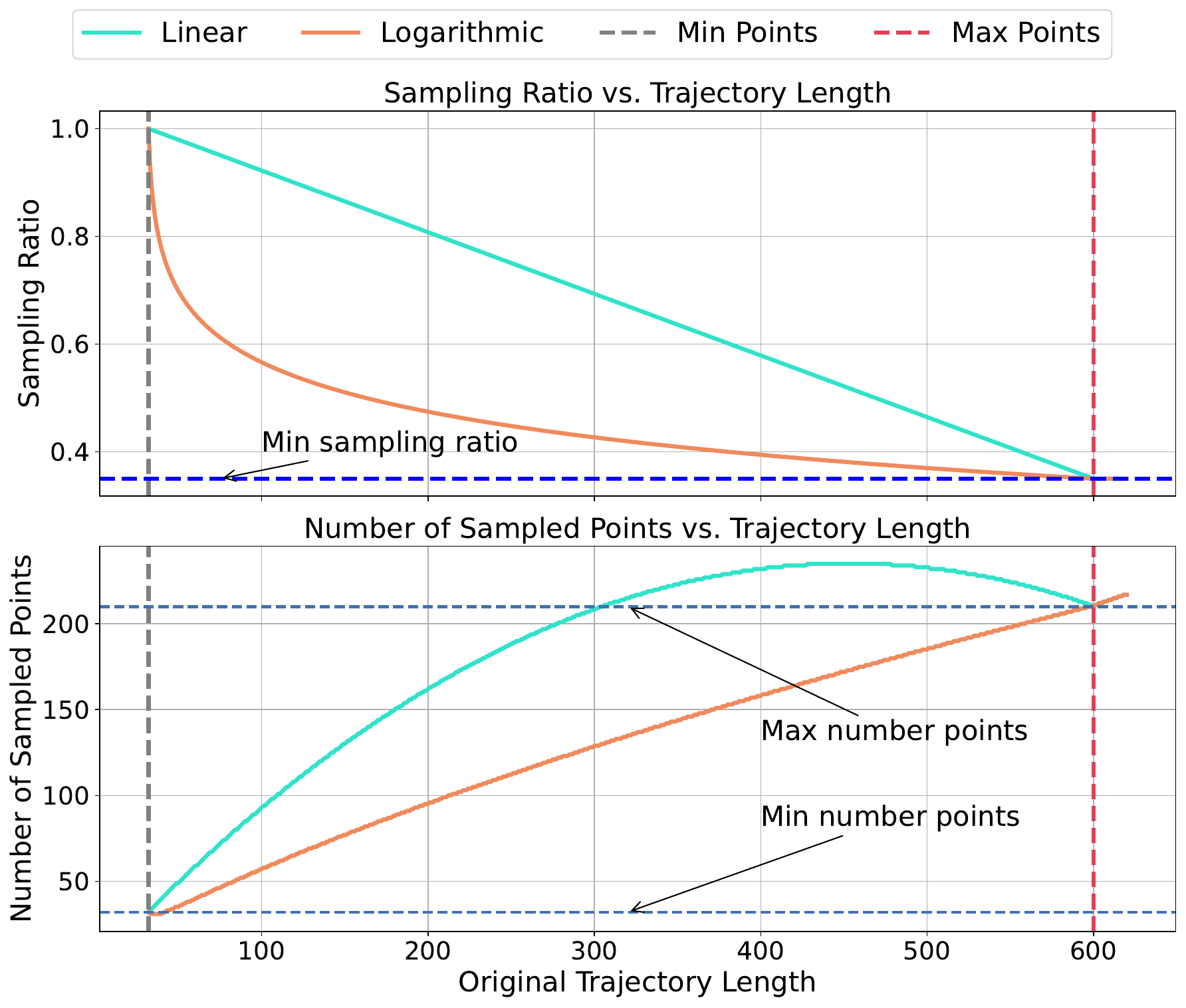}
    \caption{Illustration of the difference between dynamic resampling strategies with linear method.}
    \label{fig:app_resamping}
\end{figure}

\textbf{Visualization}. As shown in Figure \ref{fig:app_resamping},  we compare in detail the proposed dynamic resampling strategy with a linear resampling strategy (where the sampling ratio $R(n)$ decreases linearly with the length of the trajectory) regarding the sampling ratio and the sampled points.
Specifically, this figure illustrates the dynamic resampling strategy compared to a linear resampling approach. The top plot displays how the sampling ratio  $R(n)$  decreases with trajectory length  $n$. 
The dynamic strategy (orange curve) follows a logarithmic decrease, ensuring a smoother transition from retaining all points for short trajectories ($n \leq n_{\text{min}}$ ) to reducing redundancy for long trajectories ($ n \geq n_{\text{max}}$), with a minimum sampling ratio  $R_{\text{min}}$. 
In contrast, the linear resampling strategy (blue curve) decreases the sampling ratio at a constant rate.
The bottom plot shows the relationship between the number of sampled points and trajectory length for both strategies. 
The dynamic approach adjusts sampling more gradually, preserving detail for intermediate trajectories while minimizing redundancy in longer trajectories. 
However, linear sampling methods instead suffer from redundancy of sampling points due to the smoothly decreasing sampling rate.
This dynamic resampling strategy ensures a balance between data volume reduction and the retention of critical movement details. The visual comparison highlights the adaptive nature of the dynamic strategy.

\subsubsection{Interval Consistent Resampling}\label{app:ICR}
Consider different cities may exhibit drastically different sampling intervals due to: Varying data collection protocols (e.g., 1s in City A vs. 5s in City B) and technical limitations or regional preferences in tracking technologies.
This heterogeneity poses a serious challenge for developing universal trajectory models, as models trained on data from one region may fail to generalize to regions with different sampling characteristics.
Therefore, we performed consistent interval sampling (at random time intervals) on the original dataset to ensure its generalizability across different datasets.
Specifically, ICR standardizes the temporal intervals between trajectory points, transforming a trajectory $\boldsymbol{\tau} = \{(x_1, y_1, t_1 ), (x_2, y_2, t_2 ), \ldots, (x_n, y_n, t_n)\}$ with irregular time intervals into a trajectory 
$\boldsymbol{\tau^{\prime}} = \{(x_1, y_1, \Delta t ), (x_2, y_2, \Delta t ), \ldots, (x_m, y_m, \Delta t)\}$ with uniform time intervals $\Delta t = t_{i+1} - t_{i}$, for all $i \in \left [1, m-1\right ]$.
 
{\large{\textit{\textbf{Corollary 2}: Temporal Regularity for Cross-Dataset Generalization}}}

\textbf{Standpoint:} Interval consistent resampling regularizes the temporal dimension of trajectory samples, enhancing the model’s ability to generalize across datasets with heterogeneous sampling rates.

\textit{Proof:} 
Let $\mathcal{D}_1$ and $\mathcal{D}_2$ be two dataset of region with average sampling intervals $\mu^{(1)}_{\Delta t}$ and $\mu^{(2)}_{\Delta t}$.
Assume the temporal pattern recognition task can be formalized as learning a function $f_{\theta}: \boldsymbol{\tau} \rightarrow Y$ where the learned parameters $\theta$ should ideally be robust to sampling rate variations.
For trajectories with irregular sampling, the model must learn the relationship:
\begin{align}
    y=f_{\theta}((x_1, y_1, t_1 ), (x_2, y_2, t_2 ), \ldots, (x_n, y_n, t_n)).
\end{align}
This requires implicitly learning the distribution of time intervals $P(\Delta T)$, which varies across datasets.
With ICR, the learning problem becomes:
\begin{align}
y = f_\theta((x_1', y_1', t_1'), (x_2', y_2', t_2'), \dots, (x_m, y_m, t_m)),
\quad \text{with } t_{i+1}' - t_i' = \Delta t_{\text{fixed}}
\end{align}
where temporal intervals are now consistently fixed, eliminating the need to learn dataset-specific temporal distributions.

\textbf{Information Entropy Analysis}. From the entropy perspective, consider trajectories from different regions $r$ with characteristic sampling intervals $\Delta t^r$, where the distribution of intervals can be modeled as:
\begin{align}
P(\Delta t \mid r) \sim \mathcal{N}(\mu_r, \sigma_r^2),
\end{align}
where $\mathcal{N}$ is a dataset distribution with region-specific mean $\mu_r$ and variance $\sigma_r^2$.
The entropy of the joint distribution of regions (or dataset) and sampling intervals is:
\begin{align}
H(\mathcal{D}, \Delta T) = H(\mathcal{D}) + H(\Delta T \mid \mathcal{D}).
\end{align}

This high conditional entropy $H(\Delta T \mid \mathcal{D})$ creates a strong statistical correlation between regions and temporal patterns, forcing region-specific model adaptations.
Interval Consistent Resampling transforms the original trajectory $\boldsymbol{\tau}$ into $\boldsymbol{\tau}^\prime$ where $
t_{i+1}^\prime - t_i^\prime = \Delta t_{\text{fixed}} \quad \forall i \in [1, m-1]$
This transformation minimizes the conditional entropy:
\begin{align}
H(\Delta T^\prime \mid \mathcal{D}) \approx 0,
\end{align}
which effectively decoupling the temporal sampling pattern from the region.
This transformation reduces dataset-specific temporal variability, thereby bringing the conditional distributions of trajectories across datasets closer in distributional space:
\begin{align}
P(\boldsymbol{\tau}^\prime \mid \mathcal{D}_1) \approx P(\boldsymbol{\tau}^\prime \mid \mathcal{D}_2).
\end{align}
The reduction means the model sees more consistent input distributions, thus reducing the domain gap in learning.

For trajectory modeling tasks that focus on spatial patterns rather than absolute temporal dynamics, information loss is minimal when resampling preserves relative temporal order and approximate speed relationships.
For a trajectory with velocity profile $v(t) = (p_{i+1} - p_i) / (t_{i+1}-t_{i})$, the constraint:
\begin{align}
\frac{\|p_{i+1}^\prime - p_i^\prime\|}{\|p_{i+1} - p_i\|} \approx \frac{\Delta t_{\text{fixed}}}{\Delta t_i}
\end{align}
ensures that relative speed information is preserved even as absolute time intervals are normalized.

\subsection{Self-supervised Trajectory Masking}\label{app:masking}
Self-supervised Trajectory Masking (STM) forms a critical component of UniTraj's pre-training strategy, enabling the model to learn robust representations from incomplete trajectory data. While we introduced the concept in the main paper, this appendix provides a more detailed examination of the theoretical foundations, implementation details, and empirical justifications for our masking approach. 
Our Self-supervised Trajectory Masking framework implements four complementary masking strategies (as illustrated in Figure \ref{fig:app_masking}), each designed to simulate different types of real-world data incompleteness and encourage specific learning objectives:

\begin{figure}[h]
\centering
    \includegraphics[width=0.8\linewidth]{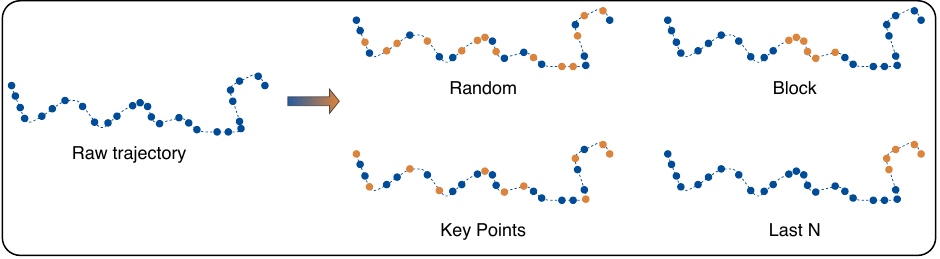}
    \caption{Illustration of the difference masking strategies.}
    \label{fig:app_masking}
\end{figure}

\subsubsection{Random Masking}
Random Masking applies a uniform probability distribution to select trajectory points for masking, where each point has an equal chance of being masked regardless of its position or significance. Formally, we select a subset of indices:
\begin{align}
\mathbf{I}_{\text{rand}} \sim \text{Uniform}(\{1,2,\ldots,n\}).
\end{align}
to mask. 
This strategy forces the model to develop both local and global dependencies, as it must learn to infer missing points from surrounding context without relying on predictable patterns. Random masking is a general masking strategy used to simulate sensor failures or temporary GPS signal loss that often occur in random trajectories.

\subsubsection{Block Masking}
Block Masking conceals consecutive segments of the trajectory by selecting a starting point $k$ and masking $b$ consecutive points: 
\begin{align}
\mathbf{I}_{\text{block}}  =\{k, k+1,\dots,k+b-1 \},\text{ for some } k.
\end{align}
This approach simulates extended sensor failures, tunnels, or urban canyons where trajectory data may be unavailable for continuous periods. The strategy challenges the model to reconstruct substantial missing segments by understanding the broader movement context, encouraging the development of long-range dependencies and trajectory continuity reasoning.

\subsubsection{Key Points Masking}

\begin{algorithm}[h]
\caption{Ramer--Douglas--Peucker (RDP) Algorithm}
\label{alg:rdp}
\begin{algorithmic}[1]
  \STATE{RDP}({$\boldsymbol{\tau}$, $s$, $e$, $\epsilon$})
    \STATE Initialize max distance $d_{\text{max}} \gets 0$
    \STATE Initialize index $k \gets -1$
    \FOR{ $i = s+1$ to $e-1$ }
      \STATE Calculate the distance from $p_i$ to $\overline{p_s p_e}$: $d_i$
      \IF{ $d_i > d_{\text{max}}$ }
        \STATE Update max distance $d_{\text{max}} \gets d_i$
        \STATE Update index $k \gets i$
      \ENDIF
    \ENDFOR
    \IF{ $d_{\text{max}} > \epsilon$ }
      \STATE $\boldsymbol{\tau}_{\text{left}}$ $\gets$ {RDP}({$\boldsymbol{\tau}$, $s$, $k$, $\epsilon$})
      \STATE $\boldsymbol{\tau}_{\text{right}}$ $\gets$ {RDP}({$\boldsymbol{\tau}$, $k$, $e$, $\epsilon$})
\RETURN $\{p_k\} \cup \boldsymbol{\tau}_{\text{left}} \cup \boldsymbol{\tau}_{\text{right}}$
    \ELSE
      \RETURN $\{p_s, p_e\}$
    \ENDIF
\end{algorithmic}
\end{algorithm}

The key points masking adopt the Ramer-Douglas-Peucker (RDP) algorithm~\cite{douglas1973algorithms}, which simplifies a trajectory by retaining points that are farthest from the line $\overline{p_1 p_n}$ connecting the first and last points. 
The indices are determined by 
\begin{align}
\mathbf{I}_{\text{key}} = \{p_k~|~d_{\text{max}}(p_{k}, \overline{p_1 p_n}) >  \epsilon \},
\end{align}
where $\epsilon$ is a predefined threshold, and $d_{\text{max}} = \max \left\{ d(p_k, \overline{p_1 p_n}) \mid 2 \leq k \leq n-1 \right\}$ is the maximum distance measures deviation from this line. 
As summarized in Algorithm~\ref{alg:rdp}, the RDP algorithm iteratively identifies the point $p_{k}$ that maximizes $d_{\text{max}} = d(p_{k}, \overline{p_1 p_n})$. If $d_{\text{max}} > \epsilon$, the corresponding point $p_k$ is treated as a key point and included in the mask set $\mathbf{I}{\text{key}}$. 
This process is recursively applied to the trajectory segments 
$\boldsymbol{\tau}_{\text{left}} = \{ p_1, \ldots, p_k \}$ and $\boldsymbol{\tau}_{\text{right}} = \{ p_k, \dots, p_n \}$, isolating critical points for masking. 
By focusing on these pivotal points, the model is challenged to reconstruct essential trajectory segments, reinforcing its understanding of key structural patterns within trajectories.

\subsubsection{Last N Masking}
Last N Masking systematically removes the final N points of each trajectory: 
\begin{align}
    \mathbf{I}_{\text{last}} =\{n-N+1, n-N+2,\ldots, n\}.
\end{align} 
This strategy explicitly simulates trajectory prediction scenarios where future positions must be forecasted based on historical observations. By incorporating this masking approach during pre-training, the model develops capabilities directly applicable to trajectory prediction tasks, creating a natural bridge between self-supervised pre-training and downstream forecasting applications.

{\large{\textit{\textbf{Corollary 3}: Robustness through Comprehensive Masking}}}

\textbf{Standpoint:} Self-supervised Trajectory Masking improves the robustness and generalization ability of the model to incomplete and heterogeneous trajectory data through a comprehensive masking strategy, enabling the model to learn more effective trajectory representations.

\textit{Proof:}
Let the trajectory data space be $\mathcal{D}$, with a true data distribution denoted as $P(\boldsymbol{\tau})$. In real-world applications, due to device limitations, communication failures, and environmental factors, the observed trajectories are often incomplete or irregular, and their distribution is denoted as $P(\tilde{\boldsymbol{\tau}})$.
The incompleteness of trajectory data can be formalized as a conditional distribution $P(\tilde{\boldsymbol{\tau}} \mid \boldsymbol{\tau})$, representing the probability of observing an incomplete $\tilde{\boldsymbol{\tau}}$ given a complete trajectory $\boldsymbol{\tau}$.

STM can be formalized as a set of masking functions $\{\mathcal{M}_1, \mathcal{M}_2, \dots, \mathcal{M}_k\}$, each corresponding to a different masking strategy. For a resampled trajectory $\boldsymbol{\tau}' = \{p_1, p_2, \dots, p_n\}$, the masking function $\mathcal{M}_i$ transforms it as:
\begin{align}
\tilde{\boldsymbol{\tau}}_i = \mathcal{M}_i(\boldsymbol{\tau}', r_i) = \{p_1, \dots, [\text{MASK}]_{j \in \mathbf{I}_i}, \dots, p_n\}
\end{align}

where $\mathbf{I} _i \subseteq \{1, 2, \dots, n\}$ is the index set of masked positions and $r_i = |\mathbf{I} _i| / n$ is the masking ratio.

\textbf{Information-Theoretic Analysis}.
From an information-theoretic perspective, STM introduces an artificial information bottleneck that forces the model to learn efficient representations. We define the model objective as minimizing the reconstruction loss:
\begin{align}
\mathcal{L}(\theta) = \mathbb{E}_{\boldsymbol{\tau} \sim P(\boldsymbol{\tau}), i \sim \mathcal{U}(1, k)} \left[d(f_\theta(\mathcal{M}_i(\boldsymbol{\tau}, r_i)), \boldsymbol{\tau})\right],
\end{align}
where $d$ is a chosen distance metric.

During training, the model needs to learn the joint distribution $P(\boldsymbol{\tau}, \tilde{\boldsymbol{\tau}}_i)$ and estimate the conditional distribution $P(\boldsymbol{\tau} \mid \tilde{\boldsymbol{\tau}}_i)$. By Bayes' theorem:
\begin{align}
P(\boldsymbol{\tau} \mid \tilde{\boldsymbol{\tau}}_i) = \frac{P(\tilde{\boldsymbol{\tau}}_i \mid \boldsymbol{\tau}) P(\boldsymbol{\tau})}{P(\tilde{\boldsymbol{\tau}}_i)}.
\end{align}
By using diverse masking strategies, the model learns to estimate $P(\boldsymbol{\tau} \mid \tilde{\boldsymbol{\tau}}_i)$ across different types of masked trajectories, which is equivalent to learning the true trajectory distribution $P(\boldsymbol{\tau})$ and the various degradation mechanisms $P(\tilde{\boldsymbol{\tau}}_i \mid \boldsymbol{\tau})$.

\textbf{Optimality Theory of Diversity Complementary Masking Strategies}.
A key innovation in STM is the use of multiple complementary masking strategies. We define the \emph{coverage region} of the union of masking strategies as:
\begin{align}
\mathcal{C}(\{\mathcal{M}_1, \dots, \mathcal{M}_k\}) = \int_{\tilde{\boldsymbol{\tau}} \in \mathcal{D}} \max_{i \in \{1, \dots, k\}} P_{\mathcal{M}_i}(\tilde{\boldsymbol{\tau}}) \, d\tilde{\boldsymbol{\tau}},
\end{align}
where $P_{\mathcal{M}_i}(\tilde{\boldsymbol{\tau}})$ denotes the distribution of incomplete trajectories generated by masking strategy $\mathcal{M}_i$.

We assert that for a suitable masking ratio and a diverse set of masking strategies $\{\mathcal{M}_1, \dots, \mathcal{M}_k\}$, the combined coverage region satisfies:
\begin{align}
\mathcal{C}(\{\mathcal{M}_1, \dots, \mathcal{M}_k\}) > \max_{i \in \{1, \dots, k\}} \mathcal{C}(\{\mathcal{M}_i\}).
\end{align}
This inequality indicates that the joint use of diverse masking functions provides strictly better coverage over possible incomplete trajectories than any individual strategy.

The advantage of combining multiple masking strategies in STM over using a single masking strategy can also be theoretically justified by comparing the expected reconstruction error.
Assume the real-world conditional distribution of incomplete trajectories is $P_{\text{real}}(\tilde{\boldsymbol{\tau}} \mid \boldsymbol{\tau})$. For a single masking strategy $\mathcal{M}_i$, let the generated distribution be $P_{\mathcal{M}_i}(\tilde{\boldsymbol{\tau}} \mid \boldsymbol{\tau})$. Then the expected reconstruction error under this distribution is:
\begin{align}
\mathbb{E}_{\boldsymbol{\tau} \sim P(\boldsymbol{\tau}), \tilde{\boldsymbol{\tau}} \sim P_{\text{real}}(\tilde{\boldsymbol{\tau}} \mid \boldsymbol{\tau})} \left[d(f_\theta(\tilde{\boldsymbol{\tau}}), \boldsymbol{\tau})\right]
\end{align}
It can be shown that training with a mixture of multiple masking strategies leads to a lower bound on this error compared to using any single strategy. This is because the mixture of diverse masking strategies better approximates the true real-world distribution of incomplete trajectories:
\begin{align}
KL\left(P_{\text{real}}(\tilde{\boldsymbol{\tau}} \mid \boldsymbol{\tau}) \,\middle\|\, \frac{1}{k} \sum_{i=1}^{k} P_{\mathcal{M}_i}(\tilde{\boldsymbol{\tau}} \mid \boldsymbol{\tau}) \right) < \min_{i \in \{1, \dots, k\}} KL\left(P_{\text{real}}(\tilde{\boldsymbol{\tau}} \mid \boldsymbol{\tau}) \,\middle\|\, P_{\mathcal{M}_i}(\tilde{\boldsymbol{\tau}} \mid \boldsymbol{\tau}) \right)
\end{align}
Here, $KL(\cdot \| \cdot)$ denotes the Kullback–Leibler divergence.
\section{Details of UniTraj}\label{app:architecture}
In this section, we provide a detailed implementation of UniTraj, including the architecture and parameter settings.

\subsection{Overall Architecture}

\begin{figure}[h]
\centering
    \includegraphics[width=0.9\linewidth]{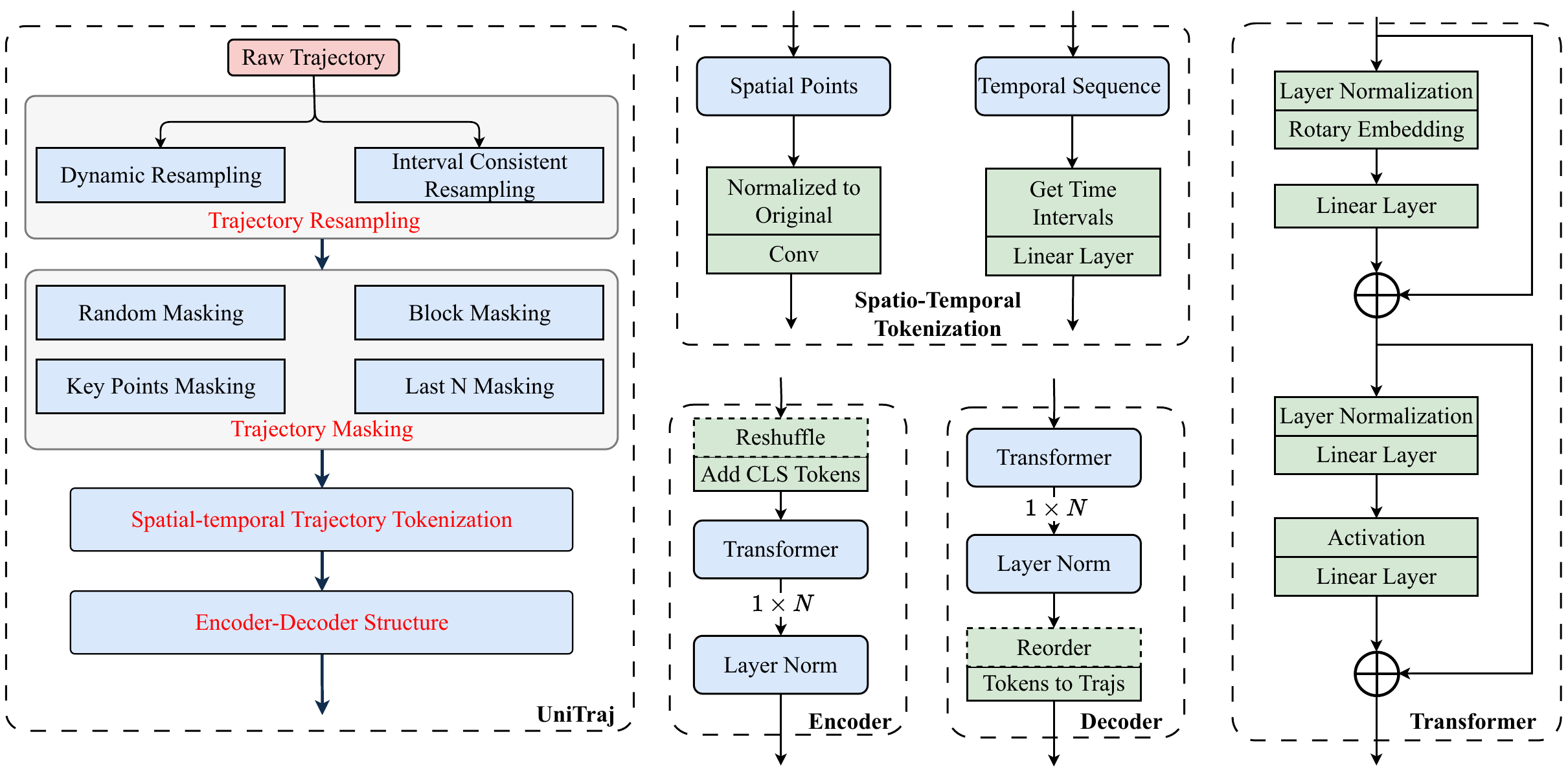}
    \caption{The main architecture and components of UniTraj.}
    \label{app_fig:unitraj}
\end{figure}

The UniTraj model adopts an encoder-decoder architecture based on transformer blocks, designed to process trajectory data with minimal regional dependency and maximum task adaptability. 
Figure \ref{app_fig:unitraj} illustrates the overall framework of UniTraj, which consists of several key components: spatio-temporal tokenization, encoder, decoder, and rotary embedding layers.

Our model takes trajectory points that have already undergone adaptive resampling (ATR) and masking (STM) as described in Appendix \ref{app:pretraining}. 
The input trajectories are represented as sequences of latitude-longitude coordinates and timestamps: 
 $\boldsymbol{\tau}=\{(\textnormal{lng}_i, \textnormal{lat}_i, t_i) | i=1,~2,\ldots,n\}$, where $n$ is the total number of points after resampling. 
 Unlike previous approaches that rely on region-specific features or road network information, UniTraj operates solely on these basic coordinates, enhancing its universal applicability across diverse geographic contexts.

\subsection{Input Representation and Embedding}

\textbf{Spatio-Temporal Tokenization}. To enhance numerical stability and generalization, all input coordinates are normalized relative to the first point in the trajectory $(x_i,y_i)= (\text{lng}_i - \text{lng}_1, \text{lat}_i - \text{lat}_1)$.
For the spatial component, we project the normalized coordinates into a $d$-dimensional space using a 1D convolutional neural network, yielding a spatial embedding 
\begin{align}
    \boldsymbol{h}_i^{s}= \text{Conv1D}([x_i, y_i]; \theta_s),
\end{align}
where $\theta_s$ represents the learnable parameters of the convolutional layer. We use a kernel size of 1 with no stride to capture local spatial dependencies.
Similarly, the temporal component, based on the time intervals $\Delta t_i$, is embedded into the same $d$-dimensional space via a linear layer, resulting in a temporal embedding:
\begin{align}
\boldsymbol{h}_i^{t} = W_t \cdot \Delta t_i + b_t,
\end{align}
where $W_t \in \mathbb{R}^{d \times 1}$ and $b_t \in \mathbb{R}^d$ are learnable parameters.
The final embedding for each trajectory point is obtained by element-wise addition of the spatial and temporal components:
\begin{align}
    \boldsymbol{h}_i = \boldsymbol{h}_i^{s} + \boldsymbol{h}_i^{t}
\end{align}
This dual-tokenization captures both spatial and temporal dynamics, enabling the model to learn relative movement and temporal dependencies effectively.

\textbf{Rotary Positional Encoding (RoPE)}. In addition to encoding the spatial and temporal details of each trajectory point, it is essential to capture the relative positional relationships between points. 
These relationships enable the model to comprehend the movement sequence and the timing between points, both crucial for accurate trajectory modeling. 
To achieve this, we employ Rotary Position Encoding (RoPE)~\cite{su2024roformer}, which maintains the relative positional information between points by rotating the trajectory embedding vectors.
Given the combined spatial-temporal embeddings $\boldsymbol{h}_i$ for point $i$ in the trajectory, RoPE applies a rotational transformation:
\begin{align}\label{eq:rope}
    \text{RoPE}(\boldsymbol{h}_i) = \begin{pmatrix}
\cos \theta_i & -\sin \theta_i  \\
\sin \theta_i & \cos \theta_i 
\end{pmatrix} \begin{pmatrix}
\boldsymbol{h}_i^{(1)} \\
\boldsymbol{h}_i^{(2)}
\end{pmatrix},
\end{align}
where $\boldsymbol{h}_i^{(1)}$ and $\boldsymbol{h}_i^{(2)}$ are the first and second halves of the embedding $\boldsymbol{h}_i$, and $\theta_i$ is a rotation angle that varies proportionally with the position index $i$.
Specifically, $\theta_i$ is calculated as $\theta_i = \frac{i}{10000^{2k/d}}$, where $k$ is the index of the embedding dimension, and $d$ is the total dimension of the embedding.

The main advantage of RoPE is its ability to preserve relative positional information through rotational symmetry. This ensures that the relative distance and directional relationships between points are maintained, enabling the model to capture both local patterns (e.g., short-term movements) and global patterns (e.g., long-range directionality) within a trajectory. By encoding these relative positions, RoPE strengthens the model’s capacity to understand movement dynamics across varying scales.

\subsection{Adaptive Representation Learning}
The UniTraj employs an encoder-decoder architecture~\cite{he2022masked} tailored for trajectory data. The encoder and decoder use Transformer blocks~\cite{vaswani2017attention} with RoPE-powered self-attention mechanisms to capture dependencies within trajectory embeddings.

\textbf{Encoder}.
Given a masked trajectory $\tilde{\boldsymbol{\tau}} = \{ p_1, \dots, \text{[MASK]}_{i \in \mathbf{I}}, \dots, p_n \}$, we first extract the embedding representations of the unmasked points $\mathbf{H} = \{ \boldsymbol{h}_1, \boldsymbol{h}_2, \dots, \boldsymbol{h}_m \}$ (where $m \leq n$ and $i \notin \mathbf{I}$) through the tokenizer and positional encoding steps. 
The encoder $\mathbf{E_{\theta}}$  processes the visible (unmasked) points in a trajectory to generate contextualized representations. It consists of $L_e$ transformer blocks, each incorporating:
\begin{enumerate}[leftmargin=*]
    \item\textbf{Multi-head Self-attention with RoPE:} As described previously, we apply RoPE to the self-attention mechanism:
    \begin{align}
    \text{Attention}(Q, K, V) = \text{softmax}\left(\frac{Q_{\text{RoPE}} \cdot K_{\text{RoPE}}^T}{\sqrt{d_k}}\right) \cdot V
    \end{align}
    where $Q_{\text{RoPE}}$ and $K_{\text{RoPE}}$ are the query and key matrices with RoPE applied.
    \item \textbf{Feed-forward Network (FFN):} A two-layer FFN with GELU activation:
    \begin{align}
    \text{FFN}(x) = W_2 \cdot \text{GELU}(W_1 \cdot x + b_1) + b_2
    \end{align}

     \item \textbf{Layer Normalization and Residual Connections:} Each sub-block is wrapped with layer normalization (Pre-LN) and residual connections:
    \begin{align}
    \mathbf{H}' &= \text{LayerNorm}(\mathbf{H} + \text{Attention}(\mathbf{H})) \\
    \mathbf{H}' &= \text{LayerNorm}(\mathbf{H}' + \text{FFN}(\mathbf{H}'))
    \end{align}
\end{enumerate}
The encoder's output is a set of hidden representations $\mathbf{H}^e = \{\boldsymbol{h}_i^e | i = 1, 2, \dots, m\}$ for the $m$ visible points.

\textbf{Decoder}.
The decoder reconstructs the masked points based on the contextualized representations from the encoder. 
It operates by combining the encoder's embeddings with mask tokens and processing them through $L_d$ transformer layers:
\begin{enumerate}[leftmargin=*]
    \item \textbf{Input Combination:} The decoder input consists of both the encoder outputs for visible points and the mask token embeddings for masked positions:
    \begin{align}
    \mathbf{H}_0^d= \text{Reorder}\left( \left\{ \begin{array}{ll}
    \boldsymbol{h}_i = \mathbf{h}^e_{j}& \text{if } i = \text{Index}(j), \; i \notin \mathbf{I} \\
    \boldsymbol{h}^\text{mask} & \text{if } i \in \mathbf{I}
    \end{array} \right\} \right),
    \end{align}
    where $\boldsymbol{h}^\text{mask}$ represents the mask token embeddings for all masked positions.

    \item \textbf{Decoder Transformer Blocks:} The combined input is processed through $L_d$ transformer blocks, each with the same structure as the encoder blocks (self-attention with RoPE, FFN, layer normalization, and residual connections). The self-attention mechanism allows information to flow between visible and masked positions:
    \begin{align}
    \mathbf{H}_l^d = \text{TransformerBlock}(\mathbf{H}_{l-1}^d)
    \end{align}
    for $l \in \{1, 2, \dots, L_d\}$.

    \item \textbf{Output Projection:} The final layer projects the decoder's representations for the masked positions back to coordinate space:
    \begin{align}
    (\hat{x}_j, \hat{y}_j) = W_o \cdot \boldsymbol{h}_{L_d, j}^d + b_o
    \end{align}
    where $j$ indexes the masked positions, and $W_o \in \mathbb{R}^{2 \times d}$ and $b_o \in \mathbb{R}^2$ are learnable parameters. These projected coordinates are then transformed back to the original coordinate system $(\hat{\text{lng}}_j,  \hat{\text{lat}}_j) = ( \hat{x}_j + \text{lng}_1,  \hat{y}_j + \text{lat}_1)$.
\end{enumerate}
UniTraj is trained using a self-supervised learning approach with a reconstruction loss function. For each trajectory, we apply our masking strategies (random, block, key points, or last N), and the model is trained to reconstruct these masked points:
\begin{align}
    \mathcal{L}= \frac{1}{|\mathbf{I}|}\sum_{j \in \mathbf{I}}\| (\hat{x}_j,\hat{y}_j) - ({x}_j,{y}_j)  \|^2_2,
\end{align}
where $\mathbf{I}$ is the set of masked positions, and $\|\cdot\|$ denotes the L2 norm.

\subsection{Task-Specific Adaptation}
For downstream applications, UniTraj can be used in two primary ways:
\begin{enumerate}[leftmargin=*]
\item \textbf{Zero-shot Transfer:} The pre-trained model's encoder can be directly applied to extract trajectory representations for various tasks without further training.
We use the pre-trained UniTraj as a backbone and attach task-specific Multi-Layer Perceptron (MLP) adapters to the output:
\begin{align}
 \mathbf{H}^{\text{final}} = \text{MLP}(\text{UniTraj}_{\text{encoder}}(\boldsymbol{\tau}))
\end{align}
The MLP adapter typically consists of 2-3 layers with non-linear activations:
\begin{align}
 \mathbf{H}^{\text{ada}} = W_2 \cdot \text{ReLU}(W_1 \cdot \mathbf{H}^e+ b_1) + b_2
\end{align}
where $\mathbf{H}^e$ is the output from the UniTraj encoder. 
\item \textbf{Fine-tuning:} Update all parameters of the backbone and adapters with specific dataset.
\end{enumerate}

For different downstream tasks, we design specific adapter architectures:

\begin{itemize}[leftmargin=*]
\item \textbf{For Trajectory Recovery/Prediction:} We can directly use UniTraj's decoder as an adapter without any additional modifications.
\item \textbf{For Trajectory Classification:} The adapter includes pooling operations followed by fully connected layers to produce class logits.
\item \textbf{For Trajectory Generation:} The adapter interfaces with generative models by providing conditioned trajectory embeddings.
\end{itemize}

\subsection{Implementation Details}\label{app:imple}
Additionally, we summarize the list of key hyperparameters and implementation-specific settings that may be used in the implementation of \model in Table \ref{tab:unitraj_para}.
Specifically, our model contains 8 encoders and 4 decoders, each using 4 heads in the attention layer.
The model has approximately 2.38 million parameters, allowing it to balance complexity and computational efficiency. We set the embedding dimension to 128 and employ RoPE to capture spatial and temporal relationships effectively. 
Our model can handle an arbitrary length of the number of trajectory points and pad it to a length of 200. 
Naturally, due to the use of rotational positional embedding, our model holds extension capability and supports a maximum length of 512.
In addition, when performing the dynamic resampling strategy, we set the minimum number of sampling points to 36 and the maximum to 600, and its minimum sampling rate is 0.35.
Finally, we provide the probability of using various masking strategies during training, which can be further adapted to the specific task as we discussed in Section \ref{sec:modelstudy} and Table \ref{tab:ablation}.

\begin{table}[h]
\small
    \caption{General parameters setting for UniTraj.}
    \centering
    \begin{tabular}{lrr} 
    \toprule
    Parameter &  Setting value & Refer range  \\ 
    \cmidrule(lr){1-3}
    Encoder Blocks & 8 & $\ge$ 2\\
    Decoder Blocks & 4 & $\ge$ 2\\
    Attention Heads & 4 & $\ge$ 1\\
    Encode Dim & 128 & 64 $\sim$ 256\\
    Parameters of Model (Millions) & 2.38  & -- \\
    Mask ratio & 0.5  & 0.25 $\sim$ 0.75\\
    Trajectory Length Padding & 200  & 36 $\sim$ 256\\
    Maximum Length Padding & 512 & -- \\
    Minimum Trajectory Points & 36 & -- \\
    Maximum Trajectory Points & 600 & -- \\
    Minimum Sampling ratio & 0.35 & -- \\
    Random Masking & 0.7  & --\\
    Key Points Masking & 0.15  & --\\
    Block Masking & 0.05  & --\\
    Last N Masking & 0.1  & --\\
    
     \bottomrule
    \end{tabular}
    \label{tab:unitraj_para}
\end{table}
\section{Experiments Details}\label{app:exp_setup}

We use the Adam optimizer and mean square error loss with an initial learning rate of $1 \times 10^{-3}$ with a learning rate scheduler. 
The model is trained for 200 epochs with a batch size of 1024, and early stopping is applied based on validation performance.
All experiments were conducted using PyTorch, where the foundation model is trained on NVIDIA A100/L40s 40GB GPUs and the baseline experiments are performed on RTX 2080 Ti.

\subsection{Datasets}\label{app:exp_dateset}
We evaluate the performance of the proposed model using six diverse real-world trajectory datasets. 
Each dataset represents different data collection scenarios, quality levels, motion patterns, and geographic regions, providing a comprehensive test of the capabilities of \model.
\begin{itemize}[leftmargin=*]
    \item \textbf{WorldTrace}: WorldTrace is our proposed large-scale, globally distributed dataset, which we describe in detail in Section \ref{sec:worldtrace}. 
    We curated a high-quality subset of 1.1 million trajectories from the original dataset, which have been filtered to remove long stops and loops. 
    Of this subset, 1 million trajectories are designated for model training combined with resampling or masking strategies, with the remaining 100,000 reserved for testing without any operation. 
    To ensure consistency and enable independent \textbf{zero-shot evaluations}, the testing dataset is normalized to a sampling interval of 3 seconds per point.    
    
    \item \textbf{Chengdu}~\cite{gaia}: 
    The Chengdu dataset comprises over one million urban mobility trajectories collected from taxis operating in Chengdu, China, reflecting daily commuting and transportation patterns in a densely urbanized area. 
    It features dense, high-frequency (3-second for most trajectories) sampling points that provide detailed insights into active urban environments.
    
    \item \textbf{Xi’an}~\cite{gaia}: 
    Similar to Chengdu, the Xi’an dataset includes millions of taxi trajectories gathered in Xi’an, China, focusing on movement patterns within another densely populated Chinese city.
    The data, collected during November 2016, captures the traffic dynamics and urban mobility behaviors specific to this region.
    
    \item \textbf{GeoLife}~\cite{zheng2009mining}: 
    The GeoLife dataset is a widely used trajectory dataset collected over three years by 182 users, primarily in Beijing, China. 
    It is mainly distinguished by a wide variety of travel modes, including walking, cycling and driving. 
    With this data, we can study the trajectory movement patterns and behavioral habits of different travel modes.
    Besides, this dataset suffers from irregular and often long sampling intervals, which limit its granularity and quality for trajectory analysis. 

    \item \textbf{Grab-Posisi}~\cite{huang2019grab}: 
    Sourced from Southeast Asia, this dataset contains 84,000 ride-hailing trajectories, predominantly from the Grab service in cities such as Jakarta and Singapore. The variable sampling intervals across these trajectories provide insights into urban mobility patterns unique to Southeast Asian metropolises.

    \item \textbf{Porto}~\cite{Portodata}:
    The Porto dataset consists of taxi trajectories collected in Porto, Portugal, capturing trips between different areas of the city.
     Although it provides valuable insight into taxi mobility within the city, the dataset has a relatively low sampling frequency, with long intervals  (15 seconds) between data points.

\end{itemize}

\subsection{Tasks Applicability Study Settings}\label{app:task_setting}

\subsubsection{\textbf{Trajectory Recovery}}
In this experiment, we randomly mask 50\% of trajectory points and test the recovery performance. 
Specifically, we evaluate \model in both \textbf{zero-shot} (trained solely on \datasetname) and \textbf{fine-tuned} settings (trained on \datasetname and then fine-tuned on each respective dataset), aiming to understand its adaptability with and without task-specific training. 
Additionally, we compare \model against a diverse range of baselines, including traditional deep learning models (Linear, DHTR~\cite{wang2019deep}, Transformer~\cite{vaswani2017attention}, and DeepMove~\cite{feng2018deepmove}) and pre-trained models (TrajBERT~\cite{si2023trajbert} and TrajFM~\cite{lin2024trajfm}). 
Performance metrics include Mean Absolute Error (MAE) and Root Mean Squared Error (RMSE) with meters, computed based on geographic distance:
\begin{align}
     {\rm MAE} &= \frac{1}{n}\sum_{i}^{n}\left|{y_{i}-\hat{y}_{i}}\right|,\\
     {\rm RMSE} &= \sqrt{\frac{1}{n}\sum_{i}^{n}\left(y_{i}-\hat{y}_{i}\right)^{2}}
\end{align}
where $y_i$ and $\hat{y}_{i}$ are the real and recovered coordinates, respectively.

\subsubsection{\textbf{Trajectory Prediction}}
In this task, we focus on predicting future trajectories based on historical trajectory points. Following the setup ~\cite{lin2024trajfm} in previous work, we predicted the locations of five future points.
The baseline settings and evaluation metrics are consistent with those used for the trajectory recovery task, and experiments were conducted on \datasetname, Chengdu, and GeoLife datasets.

\subsubsection{\textbf{Trajectory Classification}}
The Trajectory Classification task is conducted on two datasets, GeoLife and Grab-Posisi.
In this task, we will only use the encoder module of the \model as a backbone and then add a classification header.
We compare \model in two settings: without fine-tuning (wo/ft), where only the classifier head is trained, and with fine-tuning (ft), where the entire model is updated.
For baselines, we following prior literature~\cite{liang2022trajformer} use representative classification models including GRU, LSTM, STGN~\cite{zhao2020go}, and TrajFormer~\cite{liang2022trajformer}. Performance is reported by classification accuracy:
\begin{align}
     {\rm Acc} &= \frac{1}{n}\sum_{i}^{n}\mathbf{I}({y_{i},\hat{y}_{i}}),
\end{align}
where $y_i$ and $\hat{y}_{i}$ are the predicted and true labels, respectively, and $\mathbf{I}(\cdot)$ is a indicator function.
Following the general settings of previous work, we selected four travel modes from the Geolife dataset, namely walking, bus, bike, and driving.
For the Grab-Posisi dataset, there are two travel modes: car and motorcycle.

\subsubsection{\textbf{Trajectory Generation}}
In this task, we follow the approach in prior work~\cite{zhu2024controltraj}, assessing trajectory generation using sequences of road segments that represent trajectories without explicit temporal attributes.
Specifically, we use ControlTraj as a downstream task for trajectory generation, where we replace the road segment extraction component (RoadMAE) of the ControlTraj with UniTraj's encoder,  testing the effectiveness of the embedded representation. 
The evaluation includes \textbf{density error} metrics~\cite{zhu2024controltraj}:
\begin{equation}
   \text{Density Error}= \operatorname{JSD}(G\|O)=\frac{1}{2} \mathbb{D}( \| \frac{(G+O)}{2}) + \frac{1}{2} \mathbb{D}(G \| \frac{(G+O)}{2}),
\end{equation}
where $G$ is the distribution of the generated trajectories in the city (which divides each city into grids of 16×16 size and calculates the count of trajectory points associated with each grid), and $O$ is the distribution of the original trajectories.
$\operatorname{JSD}(\cdot)$ is the Jenson-Shannon divergence for two distributions.

\begin{figure*}[h]
\centering
\subfigure[Original.]{
        \includegraphics[width=0.31\linewidth]{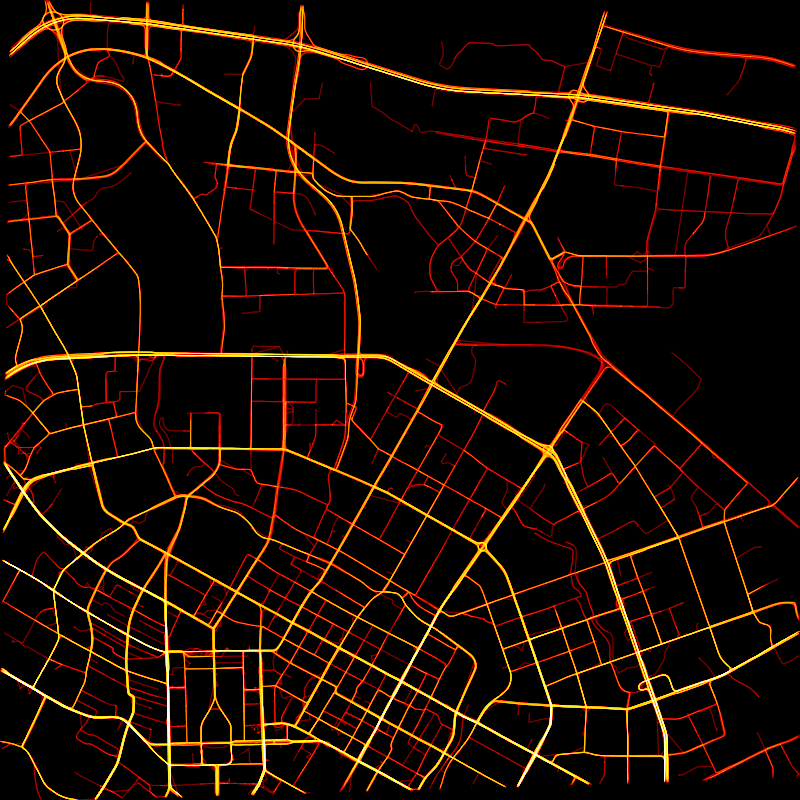}
    }  
    \subfigure[ControlTraj.]{
        \includegraphics[width=0.31\linewidth]{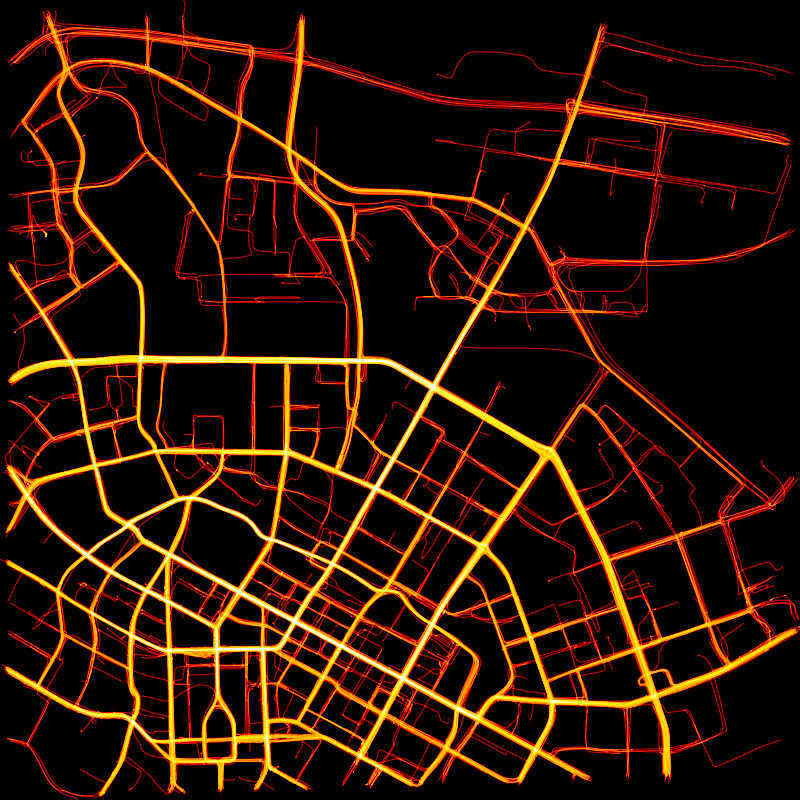}
    }
    \subfigure[ControlTraj + UniTraj.]{
        \includegraphics[width=0.31\linewidth]{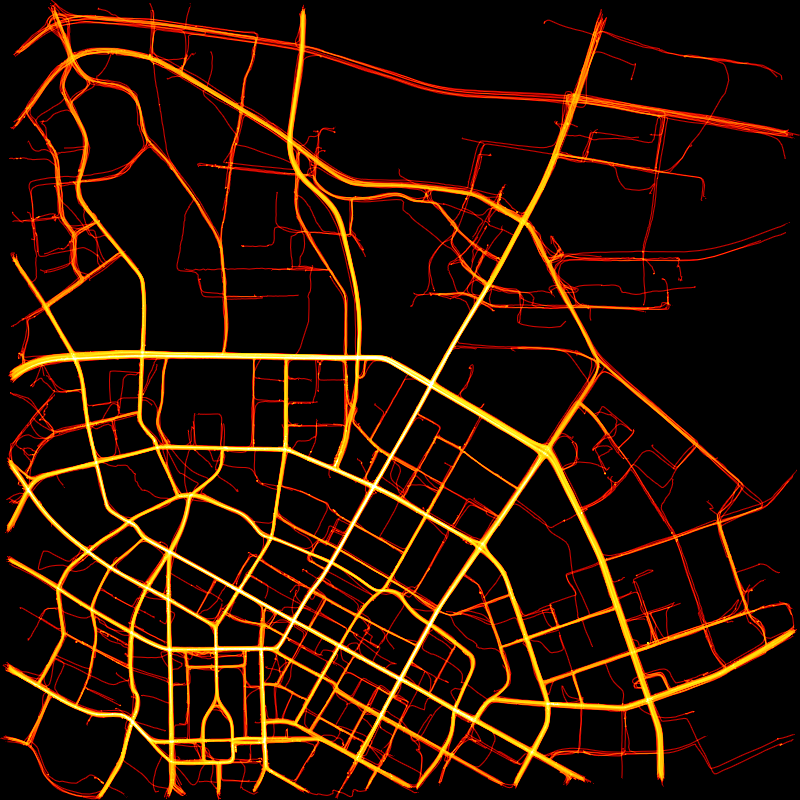}
    }\\
    \subfigure[Original.]{
        \includegraphics[width=0.31\linewidth]{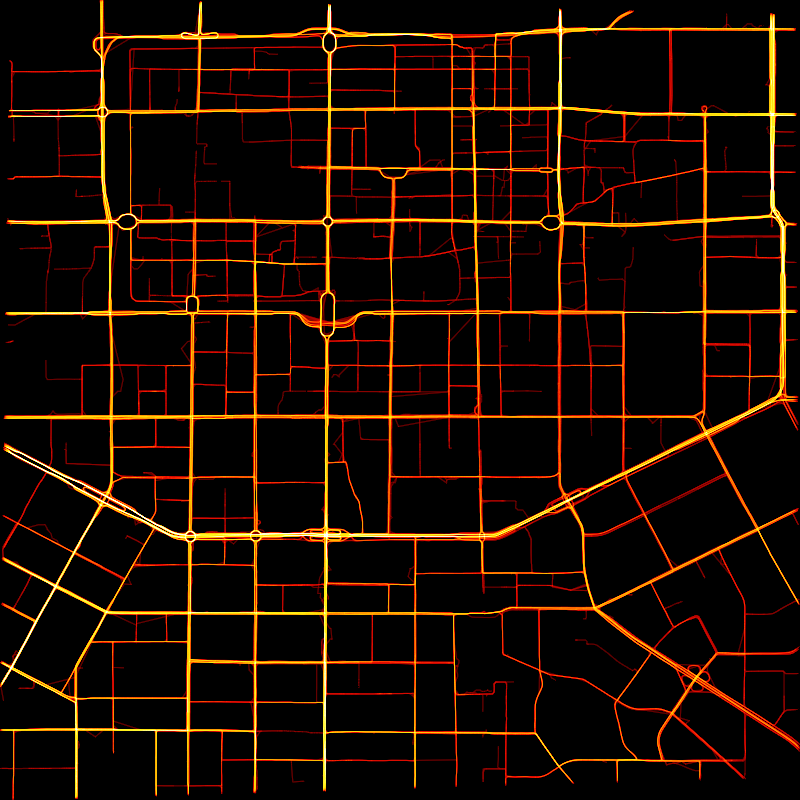}
    }
    \subfigure[ControlTraj.]{
        \includegraphics[width=0.31\linewidth]{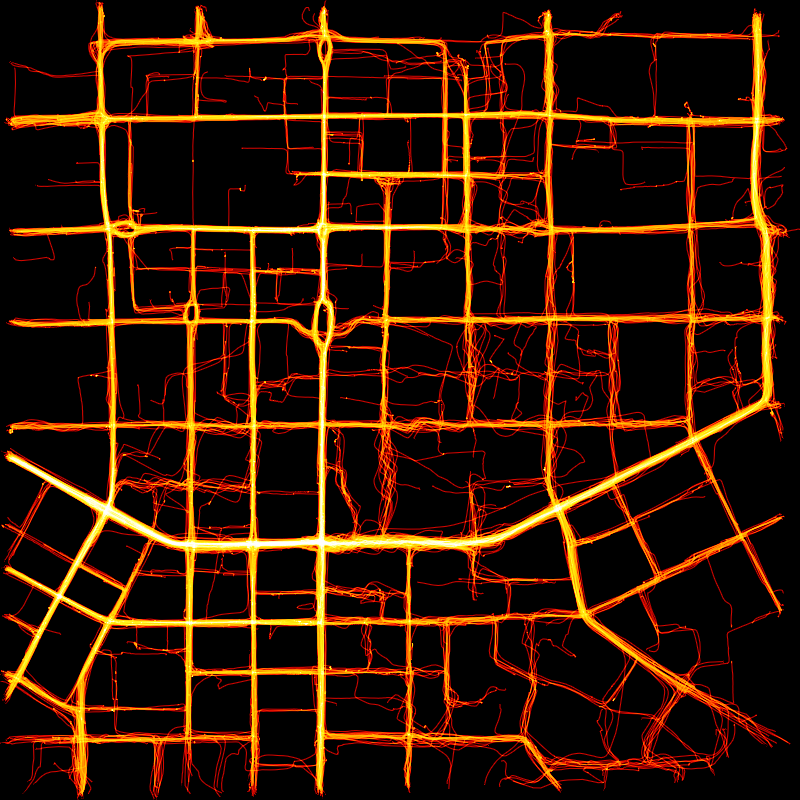}
    }
    \subfigure[ControlTraj + UniTraj.]{
        \includegraphics[width=0.31\linewidth]{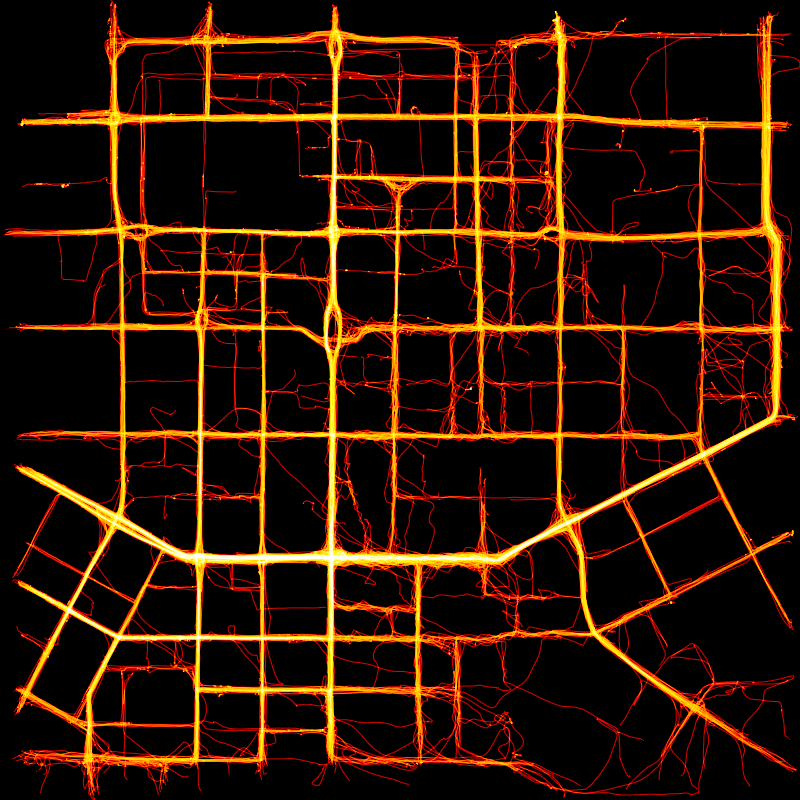}
    }
   \caption{Performance comparison of trajectory generation task with Chengdu dataset (first row), and transfer to Xi'an dataset (second row).}
   \label{app:fig_exp_generation}
\end{figure*}

For the this task, UniTraj demonstrates its versatility through integration with existing generative frameworks. By replacing ControlTraj's road segment extraction module with UniTraj, we achieved a 5.1\% reduction in density error (from 0.0039 to 0.0037) when trained and generated on the Chengdu dataset. 
This improvement, though modest in magnitude, represents a significant advance in trajectory fidelity.
More impressively, when transferring the generation capability to Xi'an without retraining---a challenging cross-region scenario---the UniTraj-enhanced generator maintains a density error of 0.0152. 
In contrast, the baseline ControlTraj experiences a 0.0171 density error when transferred across regions. 
This cross-region resilience further validates UniTraj's ability to capture universal trajectory patterns that transcend specific geographic contexts.
We also show the heatmap visualizations to measure the accuracy and realism of generated trajectories in Figure \ref{app:fig_exp_generation}, where brighter regions indicate denser trajectories and darker regions indicate sparser ones.
Detailed analysis of the generated trajectories reveals that UniTraj-enhanced generation produces more realistic speed variations, particularly in complex road segments such as intersections, sparse or dense areas. 
In summary, the above results underscore UniTraj’s potential for robust and transferable trajectory generation, proving its effectiveness in both familiar and novel geographic settings.


\subsection{Dataset Study Settings}\label{app:datastudy_setting}
\textbf{Effect of Dataset Scale and Quality.}
This task focuses on the impact of dataset size and quality on UniTraj performance.
We analyze \datasetname for the effects of different amounts and qualities of training data.
Specifically, we further process the complete \datasetname dataset by removing cyclic trajectories, removing trajectories with too many stopping points and sparse trajectories.
In total, we partitioned a subset of high-quality trajectory data numbering 1 million items, and further partitioned a subset of 10,000, 500,000 trajectory data for \model training.

\noindent \textbf{Effect of Dataset Diversity.}
The task assessed the impact of using different data coverage (i.e., geographic diversity) on the model.
We evaluate the zero-shot performance of \model trained on the \datasetname and Chengdu datasets, respectively, and tested on multiple real-world trajectory datasets. 
We chose the Chengdu dataset for comparison because it has very high data quality and has the identical me collection standards as the Xi'an dataset.

\subsection{Model Study Settings}\label{app:modelstudy_setting}
For setting the number of encoders decoders for the model, we adopt the following scheme \{encoders: 2,4,6,8,12\}, \{decoders:2,2,4,4,6\}, \{attention heads:2,2,2,4,8\}.
We believe that an asymmetric encoder-decoder architecture can significantly reduce the number of parameters while maximizing the performance of the model.
And the scaling law between the number of model parameters and the size of the data will be one of the considerations in our future research and model architecture design.

\section{More Discussion}\label{more_discuss}

\subsection{Limitation} 
While UniTraj represents a significant advancement in universal trajectory modeling, several limitations remain that warrant acknowledgment and future investigation. 
Despite WorldTrace's unprecedented geographic coverage spanning 70 countries, data distribution remains uneven, with certain regions (particularly in Africa and parts of Asia) underrepresented, potentially limiting model performance in these areas.
Additionally, our focus on motorized movement may restrict generalization to non-motorized mobility patterns, such as pedestrian trajectories with distinctly different motion properties. 
The computational resources required for training and deploying UniTraj at scale present practical challenges for resource-constrained environments, necessitating more efficient architectures or distillation approaches.
From a technical perspective, UniTraj relies solely on coordinate and temporal information, lacking integration of contextual features like road networks, traffic conditions, and points of interest that could further enhance predictive accuracy.
Addressing these limitations represents promising directions for future research, potentially through expanded geographic coverage, multimodal trajectory integration, architecture optimization, context-aware modeling, and continual learning  techniques.
Nonetheless, we believe that the proposed UniTraj and WorldTrace datasets will contribute to the development of the entire community towards a more generalized, global view of trajectory analysis.

\subsection{Broader Impact}
This work presents both promising opportunities and notable concerns for society. Positively, this universal trajectory model could popularize mobility intelligence across diverse regions, enabling improved transportation systems in underserved areas without extensive local data collection. The model could drive more efficient urban planning, reduce traffic congestion and emissions, and enhance logistics optimization globally. However, this technology could also enable more pervasive monitoring capabilities, raising surveillance concerns if misused. Additionally, there exists potential for widening technological disparities between resource-rich and resource-constrained organizations. 
Balancing these implications requires commitment to privacy-preserving techniques and equitable access policies to ensure this technology advances social welfare while minimizing potential harms.



\end{document}